\documentclass{aa}  
\usepackage{graphicx}
\usepackage{txfonts}
\usepackage{xcolor}
\begin{document}

   \title{The COSMOS Wall at $z\sim0.73$: star-forming galaxies and their evolution in different environments
    \thanks{Based on observations collected at the European Southern Observatory, Cerro Paranal, Chile, using the Very Large Telescope under program ESO 085.A-0664.
    }
   }

   \author{S.~ Zhou\inst{1},
          A.~Iovino \inst{1},
          M.~ Longhetti \inst{1},
          M.~Scodeggio \inst{2},
          S.~Bardelli\inst{3},
          M.~Bolzonella\inst{3},
          O.~Cucciati\inst{3},
          F. R. Ditrani\inst{1,4},
          A.~Finoguenov \inst{5},
          L.~Pozzetti\inst{3},
          M.~Salvato\inst{6},
          L.~Tasca  \inst{7}, 
          D.~Vergani\inst{3} 
          and
          E.~Zucca \inst{3}
          } 

   \institute{INAF-Osservatorio Astronomico di Brera, via Brera 28, I-20121 Milano, Italy\\  
    \email{shuang.zhou@inaf.it}
    \and{INAF - IASF Milano, Via Bassini 15, I-20133, Milano, Italy}  
    \and{INAF - Osservatorio di Astrofisica e Scienza dello Spazio di Bologna, via Gobetti 93/3, I-40129 Bologna, Italy}  
    \and{Università degli studi di Milano-Bicocca, Piazza della scienza, I-20125 Milano, Italy} 
    \and{Department of Physics, University of Helsinki, Gustaf Hällströmin katu 2, 00560 Helsinki, Finland}  
    \and{Max Planck Institute for Extraterrestrial Physics, Giessenbachstr. 1, 85748 Garching, Germany}  
    \and {Laboratoire d'Astrophysique de Marseille, CNRS-Universit{\'e} d'Aix-Marseille, 38 rue F. Joliot Curie, F-13388 Marseille, France}  
             }

   \date{Received XXX; accepted XXX}

  \abstract
   {}
   {We present a study of the evolution of star-forming galaxies within the so-called Wall structure at z$\sim$0.73 in the field of the COSMOS survey. We use a sample of star-forming galaxies from a comprehensive range of environments and across a wide stellar mass range. We discuss the correlation between the environment and the galaxy's internal properties, including its metallicity from the present-day gas-phase value and its past evolution as imprinted in its stellar populations.}
   {We measure emission-line fluxes from stacked spectra of galaxies selected within small stellar mass bins and in different environments. These fluxes are then converted to gas-phase metallicities. In addition, we build a simple yet comprehensive galaxy chemical evolution model, which is constrained by the gas-phase metallicities, stacked spectra and photometry of galaxies to reach a full description of the galaxies' past star formation and chemical evolution histories in different environments. Parameters derived from best-fit models provide insights into the physical process behind such evolution.}
   {We reproduce the `downsizing' formation of galaxies in both their star formation histories and chemical evolution histories at $z\sim0.73$ so that more massive galaxies tend to grow their stellar mass and become enriched in metals earlier than less massive ones. In addition, the current gas-phase metallicity of a galaxy and its past evolution correlate with the environment it inhabits. Galaxies in groups, especially massive groups that have X-ray counterparts, tend to have higher gas-phase metallicities and are enriched in metals earlier than field galaxies of similar stellar mass. Galaxies in the highest stellar mass bin and located in X-ray groups exhibit a more complex and varied chemical composition.}
   {The evolution of a galaxy, including its star formation history and chemical enrichment history, exhibits a notable dependence on the environment where the galaxy is located.  This dependence is revealed in our sample of star-forming galaxies in the Wall region at a redshift of $z\sim0.73$. Strangulation due to interactions with the group environment, leading to an early cessation of gas supply, may have driven the faster mass growth and chemical enrichment observed in group galaxies. Additionally, the removal of metal-enriched gas could play a key role in the evolution of the most massive galaxies. Alternative mechanisms other than environmental processes are also discussed.}

   \titlerunning{Star Forming galaxies in the COSMOS-Wall at  $z\sim0.73$}
   \authorrunning{S.~Zhou et al.}
   
   \maketitle

\section{Introduction}

In the current galaxy formation scenario, galaxies form inside dark matter halos \citep[e.g.][]{White1978}. But unlike the dark matter halos evolution, only governed by gravitational force and well explored with detailed N-body simulations and semi-analytical models, galaxies are complex systems regulated by various physical processes \citep{Mo1998}. Internal physical processes within individual galaxies include gas inflow and outflow, star formation and chemical enrichment, and stellar and active galactic nucleus (AGN) feedback processes. Variations of these processes can lead to complicated star formation and chemical evolution paths in galaxies, which have yet to be fully explored. In addition, a wide range of external processes can also play a role during a galaxy's evolution on scales ranging from mergers with nearby galaxies to interactions with galaxy groups, clusters and the cosmic web. Investigating secular evolution and its connection with nurture-driven evolution has long been a hot topic in galaxy evolution studies.

Among different galaxy properties, the stellar mass is found to be the key factor that drives a galaxy's internal, secular evolution \citep[e.g.][]{Gallazzi2005, Thomas2005}. Galaxies of higher stellar mass are found to grow their stellar masses earlier than less massive ones and to contain a star population formed at earlier epochs and over a shorter time-span, displaying a so-called `downsizing' trend \citep{Panter2003, Kauffmann2003, Heavens2004, Panter2007, Fontanot2009, Peng2010, Muzzin2013, Peterken2020}. These observational results well agree with the predictions of the standard $\Lambda$CDM model \citep[e.g.][]{DeLucia2006, Henriques2015}, where the stars that end up in massive galaxies have formed earlier, even if the galaxy itself may have assembled at later times. Massive galaxies are also found to be richer in metals than lower mass ones \citep[e.g.][]{Gallazzi2005, Panter2007, Thomas2010}, likely due to their deeper potential well that strengthens their ability in retaining metals against the loss during feedback processes \citep[e.g.][]{Dave2012, Lilly2013}. Such a dependence also persists when considering galaxies of similar dark matter halo masses \cite[e.g.][]{Scholz-Diaz2022, Oyarzun2022, Zhou2024legac}.
 
The evolution paths of galaxies are also, not surprisingly, found to be dependent on the environment they inhabit. It has long been known that galaxies in high-density regions such as groups and clusters are more likely to be redder, which indicates lower star formation activity, and are often associated with earlier morphological types \citep[e.g.][]{Oemler1974, Dressler1980, Postman1984}. More detailed investigations, especially with large-scale surveys such as the Sloan Digital Sky Survey (SDSS,\citealt{York2000}), have revealed that such an effect likely originates from the quenching of star formation activity of satellite galaxies in groups and clusters \citep[e.g.][]{Pasquali2010, Peng2012, Wetzel2012, Wetzel2013}. Similar investigations for galaxies at higher redshifts have shown similar environmental dependencies up to a redshift of $z\sim$3 \citep[e.g.][]{Peng2010, Darvish2016, Kawinwanichakij2017}. 

The environment also impacts the evolution of galaxies' chemical composition; however, its effects appear to be less certain. From stellar population analysis, the metallicity of a galaxy is usually derived as an average over the past generations of its stars, which can be linked to the environment it inhabits \citep[e.g.][]{Sanchez2006, Thomas2010, Pasquali2010, Zheng2017}. \cite{Sanchez2006} find that early-type galaxies in low-density environments can be more metal-rich than their counterparts in denser environments. \cite{Pasquali2010} also suggests that, at a given stellar mass, satellites are older and metal richer than their central counterparts, with a difference increasing with decreasing stellar mass. However, \cite{Zheng2017} report the opposite trend: satellite galaxies are relatively more metal-poor, especially in large-scale sheets and voids.

Regarding the gas phase, the current chemical composition of the star-forming gas in galaxies can be derived from their nebular emission line strengths. Evidence in the local Universe indicates that the gas-phase metallicities of galaxies in clusters and rich environments can be 0.05dex higher than galaxies in the field or in poorer environments \citep[e.g.][]{Ellison2009}. But opposite evidence is reported in \cite{Hughes2013}, suggesting that the stellar-gas phase mass metallicity relation is nearly invariant to the environment. At higher redshifts ($z\sim$2), observations have shown a more complex picture -- for lower mass ($M_*/{\rm M}_{\odot}<10^{10}$) galaxies, cluster galaxies tend to be more metal-rich in gas compared to field galaxies. In contrast, massive galaxies ($M_*/{\rm M}_{\odot}>10^{10}$) follow an opposite trend (see \citealt{Wang2022} and references therein). The complicated observational evidence indicates that it is still unclear when and how the environment plays a role in a galaxy's evolution, especially regarding its chemical composition.

The COSMOS-Wall dataset \citep{Iovino2016} opens a unique window to investigate this issue. This dataset focuses on a complex, large-scale structure, the so-called Wall structure, located in a narrow redshift slice around $z\sim0.73$ in the COSMOS field. Such a large-scale structure contains a rich X-ray detected cluster and several galaxy groups embedded in a filamentary structure across the survey field, and it is especially suitable for analysing the environmental effect on galaxy evolution. The spectroscopic data collected for galaxies within the Wall structure in \cite{Iovino2016} enable detailed chemical analysis from emission lines and stellar absorption features. The wealth of ancillary photometric data available for galaxies in the COSMOS field additionally serves as an auxiliary constraint. The abundant spectroscopic and photometric data makes it possible to use the `semi-analytic spectral fitting’ approach \citep{Zhou2022}, going beyond the simple averaged age/metallicity and modelling the complete mass and metallicity history of a galaxy over its lifetime. By combining the evolution of gas and stellar phases into a self-consistent framework, this approach can comprehensively describe the stellar mass growth and chemical enrichment history of galaxies. The physical parameters that characterise the timescales and the rates of gas accretion and loss derived from such an approach can be more fundamentally related to the mechanisms governing the environmental effect, helping us understand the physical process hidden behind the observed environmental dependence. In this work, combining this advanced analysis tool with the unique COSMOS-Wall dataset, we expect to shed light on how the evolution of star-forming galaxies, especially the chemical evolution in gas and stellar phases, is shaped by its stellar mass and the environment it inhabits. A complementary paper \citep{Ditrani2025} will focus on passive galaxies in the COSMOS-Wall region for a parallel analysis.

This paper is set out as follows. In Sect. \ref{sec:data}, we present the COSMOS-Wall dataset and our galaxy sample selection and data analysis process. A brief introduction to the estimate of the gas-phase metallicity of our sample galaxies, as well as the introduction and implementation of the semi-analytic spectral fitting process, is presented in Sect. \ref{sec:analysis}. We then investigate the output galaxy properties and their correlation with the environment, followed by discussions about the possible physical origin of these relations in Sect. \ref{sec:results}. Finally, our key results are summarised in Sect. \ref{sec:summary}. Throughout this work, we use a standard $\Lambda$CDM cosmology with parameters $\Omega_{\Lambda}=0.7$, $\Omega_{\rm M}=0.3$ and $H_0$=70 km s$^{-1}$ Mpc$^{-1}$, which are rounded off WMAP values \citep{Bennett2003}.  All magnitudes are always quoted in the AB system \citep{Oke1974}.

\section{Data}
\label{sec:data}
\subsection{The COSMOS-Wall dataset}
This work is based on galaxies selected from the COSMOS-Wall dataset, as initially presented in \cite{Iovino2016}. This section briefly summarises the information available for the COSMOS-Wall dataset, including spectral and photometric data and the environment's definition. 
We will then discuss the sub-sample of star-forming galaxies selected for the analysis presented in this work. 

\subsubsection{The COSMOS Wall at z$\sim$0.73}
\label{ssec:data_wall}
The Cosmological Evolution Survey (COSMOS, \citealt{Scoville2007}) is a well-known astronomical survey designed to probe the formation and evolution of galaxies. It covers a 1.4$\times$1.4 deg$^2$ field (the COSMOS field), where multi-wavelength imaging (from X-ray to radio) and spectroscopic observations have been collected using the major space and ground telescopes. Within this field, a prominent large-scale structure at $z\sim$0.7 was already detected at the beginning of the survey using only photometric data \citep{Scoville2007b, Cassata2007, Guzzo2007}. This structure is usually called the COSMOS Wall structure, and the galaxies within it are the main focus of this project. 

Among the several spectroscopic surveys that have targeted the COSMOS field \citep[e.g.][]{Lilly2007, Coil2011, Comparat2015}, two are of particular interest for this paper as they target the COSMOS Wall redshift range. The first is the zCOSMOS Bright survey \citep{Lilly2007}, using the VIMOS multi-object spectrograph, mounted on the Nasmyth focus B of ESO VLT-UT3 Melipal and the R-600 MR grism. The zCOSMOS Bright survey targeted galaxies brighter than $I_{AB} = 22.5$ within the whole area of the COSMOS field to obtain spectra for $\sim$20000 galaxies, the 20K-sample \citep{Lilly2009}. 
The second one is the Wall-Survey, which targeted galaxies within the COSMOS-Wall region, K-band selected to be brighter than $K_{AB} = 22.6$ and within the photometric redshift range $0.60 \leq z_{phot} \leq 0.86$. The Wall-Survey used the VIMOS spectrograph and the same setup as zCOSMOS, but the typical exposure times varied from 1 hour to 4 hours depending on the brightness of the targets (see \citealt{Iovino2016} for a detailed description of the observational strategy adopted).  
In both cases, the chosen VIMOS instrumental setup provided spectra covering the wavelength range 5550–9450 {\AA} with a dispersion of 2.5 {\AA} pixel$^{-1}$ and spectral resolution R$\sim$600. 
In this paper, we will use spectra coming from both surveys for a total of 1277 Wall galaxies (856 galaxies from the Wall-Survey and 421 galaxies from the 20K-sample) located within the Wall Volume, as defined by the redshift limits $0.69 \leq z_{spec} \leq 0.79$ and the RA-Dec boundaries enclosing the COSMOS-Wall structure (see the outline as displayed in Figure 1 of \citealt{Iovino2016}).  Readers are referred to \cite{Iovino2016} for more details on this sample's definition and basic properties. 

In addition to the spectroscopic data mentioned above, we will use the wealth of photometric information available in the COSMOS field for our analysis. We have matched the 1277 Wall galaxies with the COSMOS2015 catalogue \citep{Laigle2016}. This catalogue provides PSF-matched photometry for galaxies in the COSMOS field in 30 photometry bands from 0.15 to 24$\,\mu m$, including:  
\begin{itemize}
    \item new NIR (YJHK) observations from UltraVISTA-DR2 \citep{McCracken2012};
    \item optical data from 6 broad-band $(B_j,V_j,g^+,r^+,i^+,z^+)$, 12 medium band (IA427–IA827) and 2 narrow band (NB711, NB816) filters on Subaru/SuprimeCam \citep{Taniguchi2007,Taniguchi2015}, as well as $u^* $ band from CFHT/MegaCam and new Hyper-Suprime-Cam (HSC) Subaru Y-band data \citep{Miyazaki2012}; 
    \item FUV and NUV fluxes from the Galaxy Evolution Explorer (GALEX, \citealt{Martin2005});
    \item 3.6$\mu {\rm m}$, 4.5 $\mu {\rm m}$, 5.8 $\mu {\rm m}$, 8.0 $\mu {\rm m}$ and 24 $\mu {\rm m}$ data from Spitzer legacy program \citep{Sanders2007}.
\end{itemize}

In this work, we will not model the dust emission when fitting the spectra and photometry of galaxies. As dust emission may become prominent in the 24\,$\mu {\rm m}$ band \citep{Conroy2013}, we exclude this band and use all the remaining 29 photometric points to build the spectral energy distribution (SED) of each galaxy, combined with the spectral data to constrain our models.

\subsubsection{Environment for Wall galaxies}
\label{ssec:envir}
The unique interest of the Wall Volume is that it contains a complex, large-scale structure in a narrow redshift range ($0.69 \leq z_{spec} \leq 0.79$), enabling a detailed analysis of the environmental effect on a galaxy's evolution at an epoch when the Universe had roughly half today's age. The Wall Volume includes galaxies that inhabit different environments: a rich cluster, many groups, both X-ray detected and poorer ones, and lower-density regions. In this work, we adopt the group catalogue provided by \cite{Iovino2016} to assign environment information for all galaxies in our sample.  The group catalogue was obtained by applying a friend-of-friends (FOF) group detection algorithm to the Wall dataset, using an optimization strategy similar to \cite{Knobel2009, Knobel2012}. The parameters of the FOF algorithm - including the range of linking length values adopted - can be found in \cite{Iovino2016}. The group finder yields 57/26/9 groups with 3/5/10 or more members with spectroscopic observation within the Wall volume, with 34/19/6 groups residing within the narrower redshift bin 0.72$\le z \le $0.74, where the so-called Wall Structure is located. 

In addition to assigning observed galaxies into groups, \cite{Iovino2016} define corrections for the survey incompleteness using a weighting scheme. This scheme accounts for the photometric redshift success rate for target selection, the mean target sampling rate, and the spectroscopic target success rate, with an additional weight to correct for the spatial inhomogeneity of the mean target sampling rate. After accounting for all these corrections, each group is assigned a weighted number of group members, which better reflects the true richness of the group. Moreover, in \cite{Iovino2016}, the group catalogue is matched with the list of XMM-COSMOS extended sources as presented in \cite{George2011} to check for the presence of an X-ray counterpart. The match yields nine groups in the Wall Volume with X-ray detection. Performing a match with the new X-ray group catalogue in the COSMOS field \citep[][obtained using Chandra deep observation of the Cosmos Legacy Survey]{Gozaliasl2019} does not change the list of matched groups presented in \cite{Iovino2016}. Interestingly, the groups possessing X-ray counterparts are generally richer ones, with at least seven spectroscopic member galaxies in each group. Their halo masses estimated using weak lensing, are all log$(M_{200}) \gtrsim 13.5 $ (see \citealt{George2011} for more details). 

\subsection{Sample selection}
\label{ssec:Ssel}

\begin{figure}
    \centering
    \includegraphics[width=0.45\textwidth]{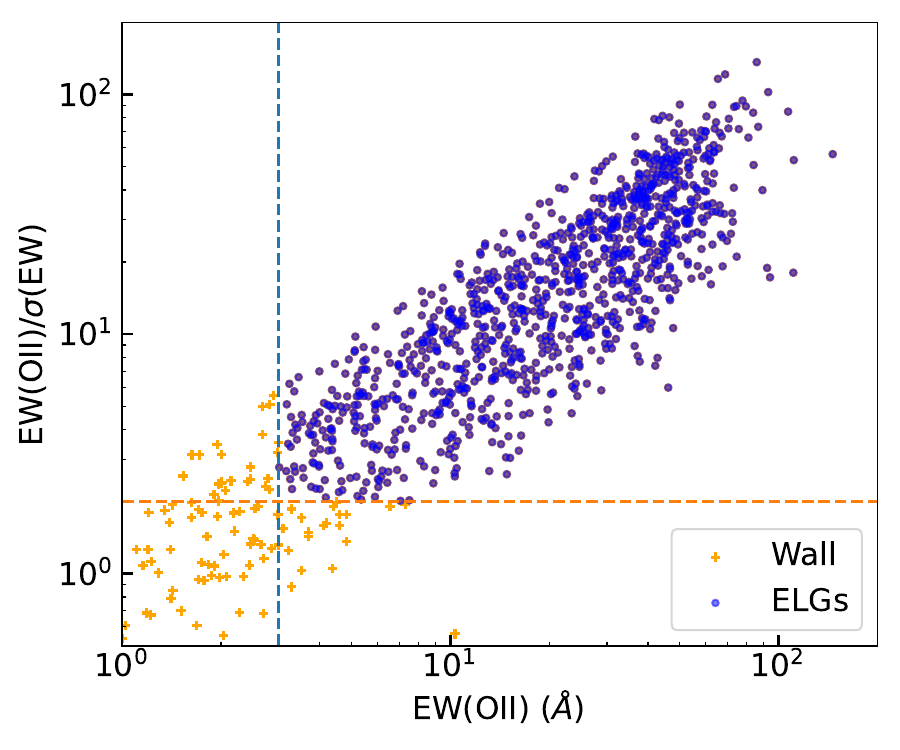}\\
     \caption{The detection significance of the \hbox{[O\,{\sc ii}]} line equivalent width, $EW(OII)/\sigma_{\rm EW}$, as a function of $EW(OII)$. Orange crosses show all the Wall galaxies, while blue dots are those identified as ELGs. The horizontal dash line marks the limit for a 2$\sigma$ detection, while the vertical dash line marks the position of $EW(OII)=3{\AA}$.}
     \label{fig:select_EW}
\end{figure}
In this section, we explain how we defined the sample of star-forming galaxies used in this study. We began by collecting all the available SED information for the complete sample of Wall Volume galaxies. The stellar masses and rest-frame magnitudes are those obtained by applying the fitting tool Hyperz-mass \citep{Bolzonella2000, Bolzonella2010} to the multi-wavelength photometric data of each galaxy, using a method similar to that outlined in \cite{Davidzon2013} (see \citealt{Iovino2016} for further details). We  carefully examined the SED shapes of the whole sample to exclude a handful of galaxies with peculiar features in their SEDs. We traced the origin of these anomalies to the blending of the galaxy flux with close companions, which are very well visible thanks to the high-quality HST optical images. 

We then moved to characterise the star-formation activity for the galaxies in the Wall Volume sample.
The Wall galaxies' spectroscopic data cover the observed wavelength range 5550{\AA} - 9650{\AA}, corresponding to the rest-frame range 3200{\AA} - 5600{\AA} at $z\sim0.73$.
The observed spectra do not cover the H$\alpha$ emission line, the most commonly used indicator of star formation activity. The observations cover the H$\beta$ line, but this line sits towards the red end of the observed spectra, where the signal-to-noise ratio is generally worse and the contamination from skyline residuals can be quite significant. We, therefore, decided to characterise the star formation activity using the \hbox{[O\,{\sc ii}]} emission, located in a more favourable spectral region. We measured for each galaxy the equivalent width of the \hbox{[O\,{\sc ii}]} emission line, i.e. $EW(OII)$, from its flux, following the procedure described in Section \ref{subsec:gas_metal_measure}. We estimated the uncertainties of such measurement ($\sigma_{\rm EW}$) using the approach discussed in Appendix \ref{app:measure}. In Fig.~\ref{fig:select_EW}, we plot the detection significance, i.e. $EW(OII)/\sigma_{\rm EW}$, as a function of the measured $EW(OII)$ for the Wall Volume sample. 
 To obtain a reliable sample of star-forming galaxies, we selected galaxies with \hbox{[O\,{\sc ii}]} emission detected at $EW(OII)>3$ {\AA} and with $EW(OII)/\sigma_{\rm EW}>2$, ensuring that only emission lines detected with at least 2 $\sigma$ significance are included in our selection \citep[an approach similar to ][]{Kong2002}. From Fig.~\ref{fig:select_EW}, we see that such a selection excludes $\sim$ 20\% of the galaxies, and we get an emission-line galaxy sample consisting of 1022 galaxies.

\begin{figure}
    \centering
    \includegraphics[width=0.45\textwidth]{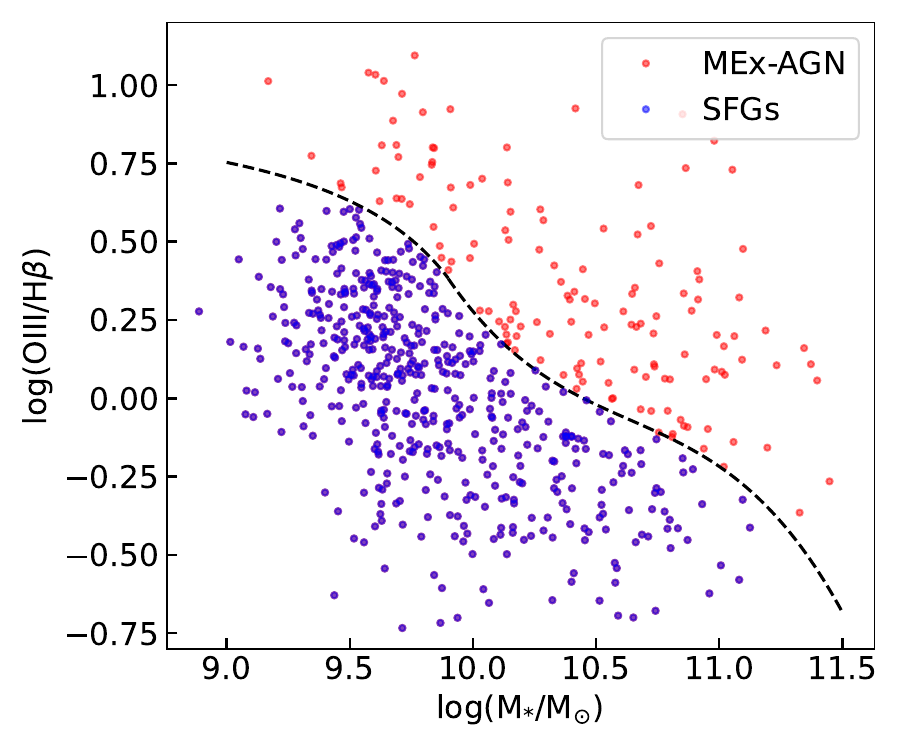}\\
     \caption{The Mass Excitation (MEx) diagnostic of ELGs in the Wall Volume sample. The dashed line indicates the separation between AGN and star-forming galaxies as proposed by \cite{Juneau2011}.}
     \label{fig:MEX}
\end{figure}

We further pruned this emission-line galaxy sample to avoid significant contamination by AGNs. Examining all the individual spectra, we first exclude a few galaxies with apparent broad emission line features, indicating strong AGN contamination. We then remove 30 X-ray detected sources using \cite{Marchesi2016}, which provides reliable multi-wavelength counterparts to the X-ray sources detected in Chandra Cosmos Legacy Survey\citep{Civano2016}. For obscured, non-X-ray detected AGNs, the limited spectral range of Wall galaxies' spectroscopic data is such that we cannot use the canonical BPT diagram to highlight the presence of AGN activity. We thus adopt the alternative Mass Excitation (MEx) diagnostic \citep{Juneau2011}, well suited to ascertain the presence of AGN contamination in the redshift range of our sample. We measured the emission line strengths of \hbox{[O\,{\sc iii}]}$\lambda$5007 and H$\beta$ lines for each of our sample galaxies using a procedure described in detail in Section \ref{subsec:gas_metal_measure}. The ratio between the obtained \hbox{[O\,{\sc iii}]} and H$\beta$ line fluxes as a function of the stellar mass of our sample galaxies is shown in Fig.~\ref{fig:MEX}. The dashed line shows the boundary separating AGN from star-forming galaxies, as proposed by \cite{Juneau2011}. Galaxies located above this line will likely be contaminated by AGN emission and thus are excluded from our sample. The final clean sample contains 924 starforming galaxies (SFGs) in the Wall Volume.

\begin{figure}
    \centering
    \includegraphics[width=0.45\textwidth]{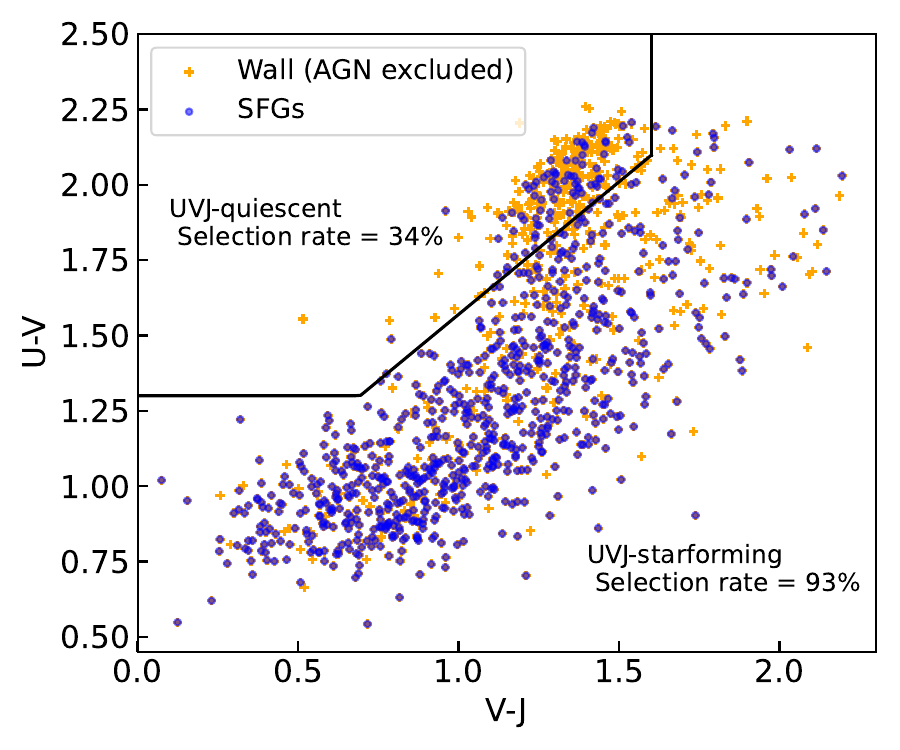}\\
     \caption{Rest frame U - V vs. V - J colours for the Wall dataset after AGN removal (orange) and SFGs selected in this work (blue). The black line separating UVJ-quiescent and UVJ-starforming regions is adopted from \cite{Whitaker2011}.}
     \label{fig:sample_UVJ}
\end{figure}

To examine the accuracy of such selection, we make use of the UVJ diagram of \cite{Whitaker2011}, which differentiates between quiescent and star-forming galaxies by using two (rest frame) colours, i.e.  (U - V) and (V - J). 
As shown in Fig.\ref{fig:sample_UVJ}, the majority of our sample comes from the star-forming region as proposed by \cite{Whitaker2011}, while only a few galaxies in the quiescent region on the UVJ diagram are also selected. This is not surprising. As already shown in \cite{Maseda2021}, more than 50\% of the galaxies at z$\sim$ 0.7 in the UVJ quiescent region have detectable \hbox{[O\,{\sc ii}]} emission ($EW(OII)>1.5 {\AA}$). However, the ionizing source of their emission is still under debate. In this work, we have applied a stricter cut with $EW(OII)>3 {\AA}$ and a 2$\sigma$ detection, and the selection rate in the UVJ-quiescent region drops to 34\%. As an additional check, we co-added the spectra of our selected SFGs in the UVJ-quiescent region and the analysis of the stacked spectrum yields a measurement of $EW(OII)=5.9 {\AA}$, confirming the presence of a non-negligible \hbox{[O\,{\sc ii}]} emission in these galaxies. Similarly, our selection rate in the UVJ starforming region is 93\%, which means that some of the galaxies that sit in the star-forming region in the UVJ diagram are excluded from our analysis. By analysing the stacked spectrum of these excluded galaxies, we confirm that they do not show detectable emission in the \hbox{[O\,{\sc ii}]} line. We can find residual H$\beta$ line emission in the stacked spectrum of these galaxies, but only with a tiny equivalent width ($\sim1.5{\AA}$), again indicating that their star formation activity (if present) is very weak. Assessing whether EW(OII) or UVJ colours are a more efficient predictor of star formation goes beyond the scope of this work. However, our analysis confirms that star-forming galaxies dominate our sample and that the contamination of non-OII-emitting galaxies is negligible. 
Therefore, we will adopt the \hbox{[O\,{\sc ii}]}-based selection throughout this work. In what follows, the selected 924 galaxies within the Wall Volume will be simply called star-forming galaxies, without further mentioning that this selection is based on their \hbox{[O\,{\sc ii}]} emission.

\begin{figure}
    \centering
    \includegraphics[width=0.5\textwidth]{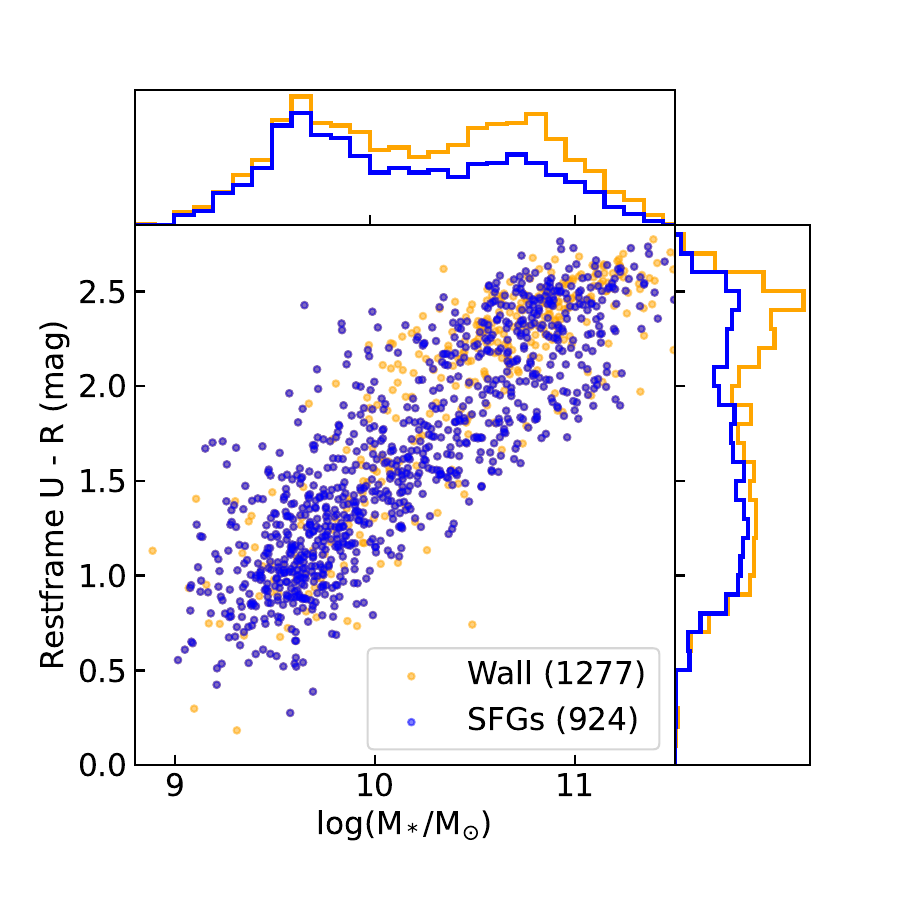}\\
     \caption{Restframe U-R colour as a function of stellar mass for the entire Wall dataset (orange) and the SFGs selected in this work (blue). Histograms on the top and right sides of the plot are marginal distributions of the stellar mass and U-R colour, respectively. }
     \label{fig:sample}
\end{figure}

Summarising the results of our selection process, in Fig.~\ref{fig:sample} we plot the rest frame U-R colour as a function of galaxy stellar mass, for the entire set of 1277 Wall galaxies (orange) and for the 924 SFGs as selected in this paper (blue). Compared to the full 1277 Wall Volume galaxies, complete down to $M_{*}/{\rm M}_{\odot}\sim10^{9.8}$ for the redder population \citep{Iovino2016}, our selection process excludes preferentially redder and more massive galaxies. As mentioned above, the contamination by non-emission line galaxies, preferentially located in high-density regions, is negligible. In our analysis, we will use the whole available stellar mass range of the SFGs sample, down to $M_{*}/{\rm M}_{\odot}\sim10^{9.0}$, any incompleteness in mass being lower for bluer galaxies and with a negligible (if any) dependence from the environment.  We can thus use our selected sample of SFGs to reliably investigate the environmental dependence of SFG's properties.

 Using the environmental information as presented in Section \ref{ssec:envir}, out of the 924 selected galaxies, we identify 259 galaxies that reside in groups with an incompleteness-corrected number of group members greater or equal than 3 (hereafter group galaxies) and 665 relatively isolated galaxies (hereafter field galaxies). In addition, out of the 259 group galaxies, 85 galaxies reside in the groups with an X-ray extended source counterpart, while the remaining 174 galaxies are in groups without an X-ray counterpart. In what follows, we will call them `X-ray group galaxies' and `non-X-ray group galaxies', respectively.  

\subsection{Data reduction and stacking}

The spectroscopic Wall data consists of spectra with median continuum signal-to-noise (S/N) values $\sim$3.5 per 1.5 {\AA} pixel in the restframe, which is not sufficient for a detailed stellar population analysis.  Thus, we choose to produce stacked spectra of galaxies to obtain spectra that have enough S/N for a reliable analysis. As the stellar mass is thought to be the most prominent factor that affects a galaxy's evolution, we first divided our galaxies into different stellar mass bins. 
 We decided to discard the stellar mass bin $M_*/{\rm M}_{\odot}>10^{11}$, containing 78 galaxies, 14 located in the X-ray group, 21 in the non-X-ray group, and 43 in the field. This subsample is not only relatively small in size, but is also predominantly composed of galaxies with lower emission line flux values. As a consequence, even after stacking, the estimates of the fluxes of \hbox{[O\,{\sc iii}]}$\lambda$5007 and H$\beta$, both essential for determining gas-phase metallicities (see Section \ref{subsec:gas_metal_measure}), are extremely noisy, and the subsequent analysis quite uncertain. Consequently, we ignored this mass bin from our analysis.  The remaining  846 galaxies are then divided into 4 stellar mass bins from $10^9{\rm M}_{\odot}$ to $10^{11} {\rm M}_{\odot}$ using a fixed bin width of 0.5 dex, and the numbers in each environment are those listed in Table \ref{tab:bins}.

In each stellar mass bin, we produced five stacked spectra for different purposes, including: 
\begin{itemize}
    \item stacking of all galaxies in each stellar mass bin as a reference (labelled `all' in the subsequent plots);
    \item stacking of `group' and `field' galaxies, respectively, in each stellar mass bin to investigate the environment dependence of galaxy properties;
    \item additional stacking of `X-ray group' galaxies and `non-X-ray group' galaxies in each stellar mass bin to explore the potential dependence of galaxies' evolution on the group's X-ray properties.
    
\end{itemize}

\begin{table}
	\centering
	\caption{Number of galaxies in each bin}
	\label{tab:bins}
        \setlength{\tabcolsep}{4.2pt}
	\begin{tabular}{lccccc}
		\hline
		Mass range & All & Group & Field& \begin{tabular}{c}
		     X-ray\\
		     Group 
		\end{tabular}  &\begin{tabular}{c}
		     Non-X-ray\\
		     Group 
		\end{tabular}\\
		\hline
		$[10^{9.0},10^{9.5}]$ & 110 & 28& 82& 11&17\\
		$[10^{9.5},10^{10.0}]$ & 329 & 78& 251& 21&57\\
		$[10^{10.0},10^{10.5}]$ & 190 & 47& 143& 14&33\\
		$[10^{10.5},10^{11.0}]$ & 217 & 71& 146& 25&46\\
  
		\hline
	\end{tabular}
\end{table}

\begin{figure}
    \centering
    \includegraphics[width=0.5\textwidth]{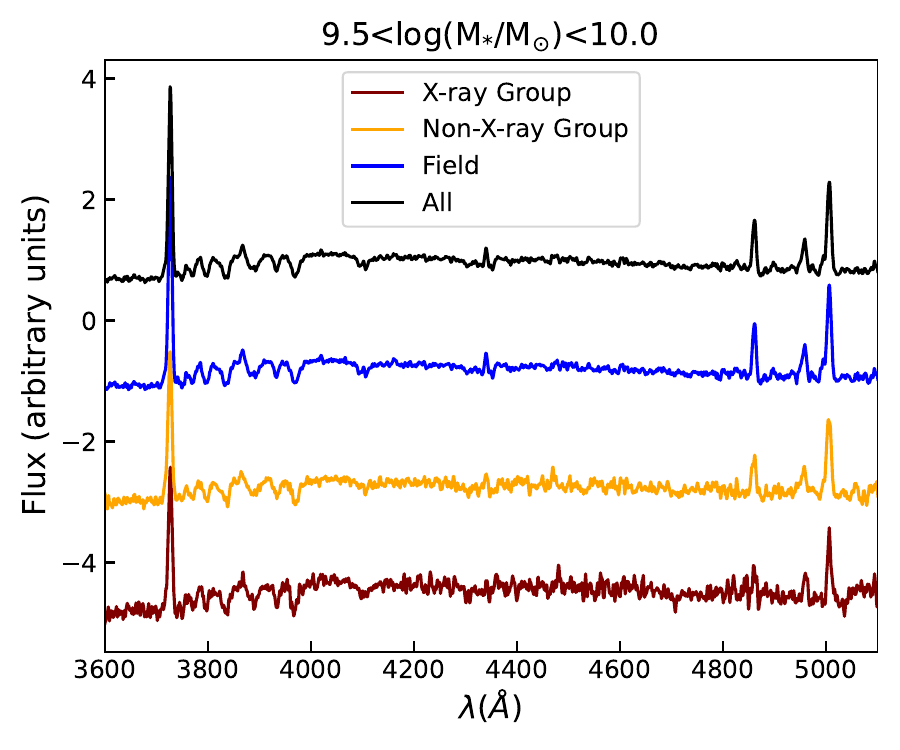}\\
     \caption{Examples of stacked spectra in the stellar mass bin $10^{9.5}<M_{*}/{\rm M}_{\odot}<10^{10.0}$. The stacking of all galaxies and of galaxies in different environments are shown with different colours, as indicated by the label. Spectra are shifted vertically to avoid overlapping.}
     \label{fig:example_spec}
\end{figure}

To perform the stacking, we first convert each galaxy spectrum to the rest frame using the spectroscopic redshift as provided by \cite{Iovino2016}. 
 The stacked spectra are then obtained as the mean of the spectra in each category listed in table \ref{tab:bins}. Such stacking produces spectra with typical estimated S/N$\sim$20 per 1.5 {\AA} pixel. In Fig.\ref{fig:example_spec}, we show as an example the stacked spectra for galaxies in the stellar mass bin $10^{9.5}<M_{*}/{\rm M}_{\odot}<10^{10.0}$ and in different kinds of environments. Although referring to galaxies in the same stellar mass bin, the plot's spectra display considerable differences in continuum shape, absorption features, and emission line strengths, potentially indicating varying evolution paths for galaxies in different environments. We also perform stacking of the photometric data presented in Section \ref{ssec:data_wall} of our sample galaxies to obtain a corresponding stacked SED for each stacked spectrum. The SED stacking is performed similarly to spectrum stacking: individual SEDs are corrected to the restframe, and the stacked SED is obtained as the mean in each bin considered.
  The broad wavelength coverage and high precision SED data greatly help constrain the chemical evolution models used in this work \citep[see][for a detailed discussion on this point]{Zhou2024legac}. In what follows, we will perform a detailed analysis of the stacked spectra and SEDs to investigate the different evolutionary paths for galaxies in different mass bins and environments. 

\section{Analysis}
\label{sec:analysis}

For each galaxy, its spectrum and SED contain information on its star formation and chemical evolution. 
The emission lines strengths and ratios encode information on the galaxy's current star formation activity and its gas-phase metallicity. On the other hand, the absorption line spectra provide evidence of the galaxy's past evolution, as imprinted in generations of stars formed at different epochs during the galaxy's lifetime. This section details the methods adopted to extract this info from the sample of coadded galaxy spectra and SEDs discussed in the previous section.

\subsection{Gas-phase metallicities}
\label{subsec:gas_metal_measure}
\begin{figure}
    \centering
    \includegraphics[width=0.5\textwidth]{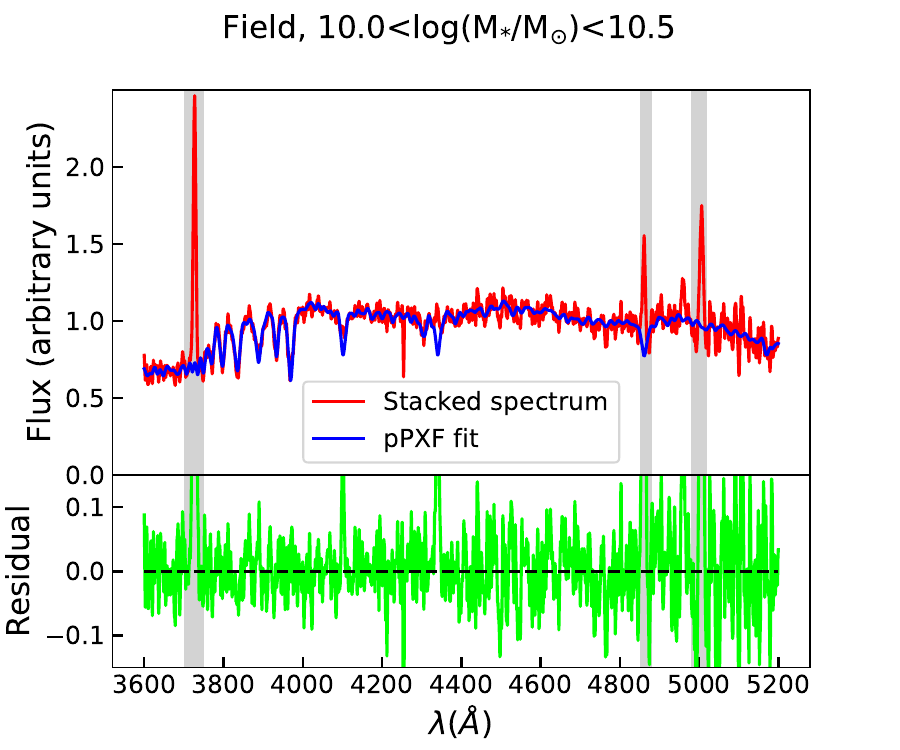}\\
    \includegraphics[width=0.47\textwidth]{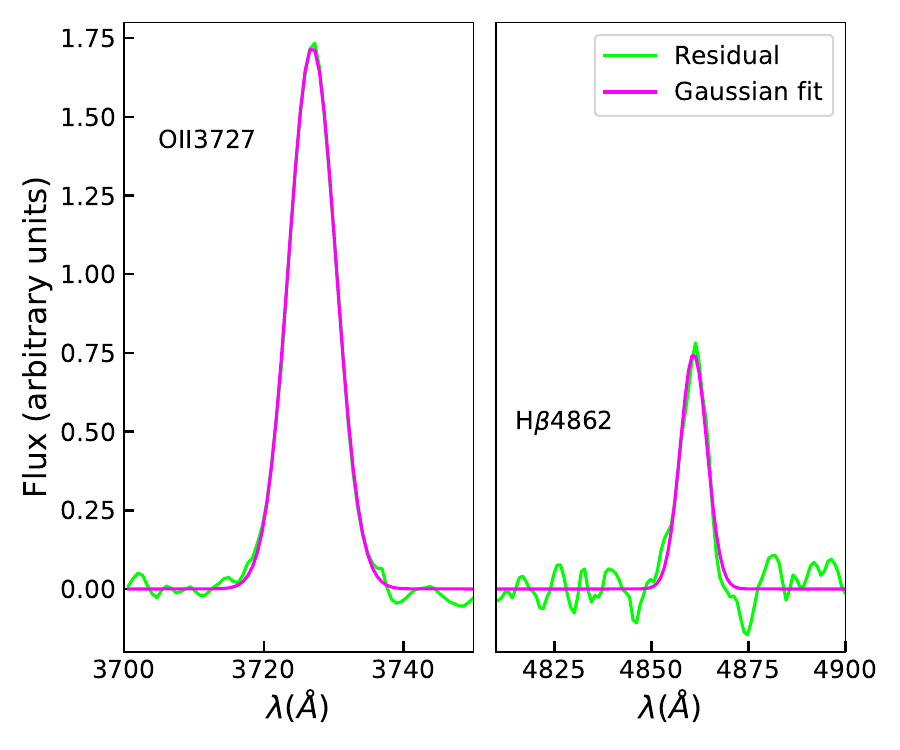}\\
     \caption{Example of this work's emission line measurement process. The top panel shows the pPXF fit (blue) to the continuum shape of the stacked spectrum (red), with residuals shown in lime green. In the bottom panel, we zoom in to the regions with restframe wavelengths around the \hbox{[O\,{\sc ii}]}$\lambda$3727 (left) and H$\beta$ (right) emission lines. In the panel, the residual spectrum is shown in lime green, with magenta lines showing our Gaussian fit to the emission lines as indicated.}
     \label{fig:example}
\end{figure}

The gas phase metallicities for the stacked spectra can be derived from emission line properties, as different line ratios indicate different ionization states and different gas phase chemical compositions \citep{Kewley2001}. To measure the emission line strength from each stacked spectrum, we first use the software pPXF \citep{Cappellari2004, Cappellari2017} to model the continuum shape. pPXF fits the continuum using SSP models from \cite{BC03}, adopting the STILIB empirical stellar spectra templates \citep{Borgne2003} and a Chabrier IMF \citep{Chabrier2003}. Additive Legendre polynomials are included to account for large-scale fluctuations in the continuum shape due to dust attenuation and possible flux calibration issues. No regularization is applied in the pPXF fitting to ensure maximum matching to the continuum. An example of the fitting output is shown in the top panel of Fig.\ref{fig:example}.

The best-fit continuum obtained from pPXF is then subtracted from the stacked spectrum to obtain a residual spectrum. We apply a single Gaussian fit to the continuum-subtracted spectrum to model the emission line profile around the  \hbox{[O\,{\sc iii}]}$\lambda$5007, \hbox{[O\,{\sc ii}]}$\lambda$3727 and H$\beta$  emission lines. As an example, in the bottom panel of Fig.\ref{fig:example}, we zoom in on the residual spectrum around the \hbox{[O\,{\sc ii}]}$\lambda$3727 and H$\beta$ lines, and our best-fit Gaussian profile to these lines are shown in magenta. 

 We obtain emission line fluxes from the best-fit model of each line profile. 
 Before converting these line fluxes into gas-phase metallicities, corrections for dust attenuation must be applied. The limited wavelength coverage of the Wall Volume sample spectra prevents us from using the Balmer decrement, i.e. the ratio of H$\beta$ and H$\alpha$ lines, to estimate dust attenuation, as  H$\alpha$ is not available for our sample galaxies. Instead, we estimate gas-phase attenuation by scaling the stellar component’s attenuation by the widely used factor E(B-V)$_{g}$=E(B-V)$_{*}$/0.44 \citep[e.g.][]{Li2021}. The stellar dust attenuation is derived using an independent pPXF fit where the additive Legendre polynomials are replaced by a \cite{Calzetti2000} dust attenuation curve.

After correcting line emission fluxes for dust attenuation, we use the calibration developed by \citealt{Kobulnicky2004} (hereafter KK04) to convert the emission line ratios into gas-phase metallicities for our sample galaxies. This approach is based on the three emission lines mentioned above (\hbox{[O\,{\sc iii}]}$\lambda$5007, \hbox{[O\,{\sc ii}]}$\lambda$3727 and H$\beta$) calibrated using stellar evolution and photoionization grids from \cite{Kewley2002} to determine the oxygen abundance in the gas phase. Note that this calibration has two branches (`lower' and `upper'), depending on the ionization state of the gas. The `lower branch' is valid for metal-poor galaxies with log(O/H)$\lesssim$8.4. According to  \cite{Zahid2011}, at z$\sim$0.7, galaxies of stellar mass higher than $10^{9.0}$M$_{\odot}$ are likely to have log(O/H)$\gtrsim$8.4, and therefore all our sample galaxies should lie on the `upper branch'. This assumption is confirmed by the results we obtain in, e.g. Fig.\ref{fig:gasmetal}, and we will use the formula for this branch to estimate the gas-phase metallicity throughout this paper. We note that, as pointed out by \cite{Kewley2008}, different calibrations of gas-phase metallicities have systematic offsets between each other. 
Throughout this work when making comparisons to literature results, we converted estimates from any other calibrations to the KK04 calibration using the formula from \cite{Kewley2008}.

\subsection{Stellar population properties}
\label{subsec:stellarpopanalysis}
We use a semi-analytic spectral fitting approach to obtain the stellar population properties and gas infall and outflow histories from our galaxy sample's stacked spectra and SEDs.
Readers are referred to \cite{Zhou2022} for details on the adopted semi-analytic fitting approach and to \cite{Zhou2024legac} for the combined fitting of spectrum and SED. Here, we briefly introduce the rationale of our fitting approach and a short description of the basic yet general chemical evolution model adopted in our fitting. 

Our model includes simple prescriptions that account for gas inflow, outflow, as well as star formation, well suited to investigate the evolution of merger-free star-forming systems. The model is fitted directly to the galaxies' absorption-line spectra, while the galaxies' emission lines constrain the current gas-phase metallicity and star formation rate. To allow a direct comparison between gas and stellar properties, we assume a solar metallicity of 0.02 and corresponding abundance patterns with log(O/H)=8.93 from \cite{Anders1989}, consistent with the Padova1994 isochrones used in \citet{BC03} SSP models. When comparing with other values of solar chemical compositions one needs to apply the appropriate transformations. 
More complex semi-analytic models (e.g., L-Galaxies; \citealt{Henriques2020}) incorporate up-to-date physical descriptions of various processes governing galaxy evolution. However, since we are working with co-added spectra rather than individual targets, it is challenging to fully constrain all these processes, and a general, albeit more coarse, modelling is satisfactory for our analysis. 

The main ingredients of our model are: 
\begin{itemize}

\item{ the time-dependent gas infall, modelled as a single exponentially decaying function with a starting time of gas infall $t_0$ and a characteristic timescale $\tau$:  
\begin{equation}
\dot{M}_{\rm in}(t)=A e^{-(t-t_0)/\tau}\ (t > t_0)
\end{equation}}

\item{the time-dependent fraction of gas turning into stars during the star formation process, as described by a linear Schmidt law \citep{Schmidt1959}: 
\begin{equation}
 \label{eq:schmidtlaw}
 \psi(t)=S\times M_{\rm g}(t) 
 \end{equation}
where $S$ is the star-formation efficiency, assumed to be $S=1~{\rm Gyr}^{-1}$ \citep{Spitoni2017, Zhou2022} while $M_{\rm g}$ is the galaxy's gas content;}

\item{the return of gas from dying stars to the interstellar medium, assumed to occur instantaneously upon star formation and with a constant mass return fraction of $R=0.3$ \citep{Spitoni2017, Zhou2022}: 
\begin{equation}
\dot{M}_{\rm ret}(t)= R\times \psi(t)
\end{equation}}

\item{the gas removal processes, whose strength is assumed to be proportional to the star formation activity, characterised by a dimensionless coefficient $\lambda$,  known as the wind parameter or mass-loading factor: 
\begin{equation}
\dot{M}_{\rm out}(t)=\lambda\times\psi(t).
\end{equation}}

\end{itemize}

The fundamental equation describing the evolution of the galaxy gas mass can thus be written as: 
\begin{equation}
\label{eq:massevo}
\dot{M}_{\rm g}(t)=
   A e^{-(t-t_0)/\tau}-S(1-R+\lambda) M_{\rm g}(t).
\end{equation}
where each term refers to one of the four ingredients detailed above. 
We further describe the gas chemical evolution as: 
\begin{equation}
\label{eq:cheevo}
\begin{aligned}
\dot{M}_{Z}(t)=& Z_{\rm in}\dot{M}_{\rm in}(t)-Z_{\rm g}(t)(1-R)SM_{\rm g}(t) + y_Z(1-R)SM_{\rm g}(t)\\
& -Z_{\rm g}(t)\lambda SM_{\rm g}(t).
\end{aligned}
\end{equation}

where $Z_{\rm g}(t)$ is the gas-phase metallicity at a function of time, so that $M_{Z}(t)\equiv M_{\rm g}\times Z_{\rm g}$ is the total mass of metals contained in the gas phase. The first term is the inflow term, and since it is generally assumed to have pristine gas infall,  we adopt $Z_{\rm in}=0$ throughout this work. The second term characterises the mass of gas locked up in low-mass stars, which are long-lived and will not return mass to the interstellar medium (ISM). The third term, in contrast, represents the chemical-enriched gas returned from dying stars. The parameter $y_Z$ is the so-called metal yield, i.e. the fraction of metal mass generated per stellar mass.
Throughout this work, we adopt $y_Z=0.063$ from \cite{Spitoni2017}, which is calculated through stellar yields form \cite{Romano2010} with assuming a Chabrier IMF. Finally, the last term represents the removal of metal-enriched gas, commonly referred to as the `outflow' term.

The picture proposed by this model is that gas infall brings material to the centre of dark matter halos, triggering star formation activity.
While massive stars die quickly and return chemically enriched gas to the ISM, gas removal processes — driven either by internal feedback from supernovae/AGN or by external mechanisms like gas stripping — deplete enriched gas from the galaxy, thereby suppressing star formation and halting further chemical enrichment. 

We are fully aware that this is a very simplified description, with different approximations that are discussed in detail in Appendix \ref{subsec:uncertainties}.
Simplified descriptions of galaxy evolution like this have been successfully applied in previous studies, including investigations of the mass-metallicity relation in star-forming and passive galaxies \citep{Spitoni2017,Lian2018mzr}, the low metallicity of passive galaxies at $z\sim0.7$\ \citep{Beverage2021}, and metallicity gradients in galaxies \citep{Belfiore2019bathtub}.

An important point is that this model does not explicitly include environmental processes such as tidal stripping \citep{Gunn1972} and strangulation \citep{Balogh2000}. However, their effects can still be inferred, albeit indirectly, from the parameters of the best fit. Tidal stripping can efficiently remove gas from a galaxy, behaving similarly to a strong outflow, while strangulation may manifest as a shortened gas infall timescale. We will argue more about this point in Section \ref{subsec:origin}, now we move to briefly present how the comparison between templates and observations is performed.

We generate a set of spectral templates by means of different combinations of the parameters from 
 a prior distribution, as listed in Table \ref{tab:paras}. Following the chemical evolution prescriptions (equations \eqref{eq:massevo} and \eqref{eq:cheevo}),
  we derived the corresponding star formation histories (SFH) and chemical evolution histories (ChEH). The SFHs and ChEHs are then combined with single stellar population (SSP) spectrophotometric models to generate the corresponding spectral templates and SEDs of composite stellar populations (CSPs) following a standard stellar population synthesis procedure (see \citealt{Conroy2013} for a review). We use SSP models from \cite{BC03}, constructed using the STELIB empirical stellar spectra library \citep{Borgne2003} and the ‘Padova1994’ stellar evolution tracks \citep{Bertelli1994} assuming the Chabrier IMF \citep{Chabrier2003}. These models provide high resolution (3{\AA} FWHM) SSP spectra in the wavelength range of 3200–-9500{\AA}, covering metallicities from Z = 0.0001 to Z = 0.05, and ages from 0.0001 Gyr to 20 Gyr, well suited for our analysis.
The templates obtained this way are then convolved with the effective velocity dispersion of each stacked spectrum as provided by pPXF (see previous section), while 
a simple screen dust model with a \cite{Calzetti2000} attenuation curve is applied to account for the dust reddening effect. To obtain the corresponding template SEDs, each template spectrum is convolved with each filter's response function. Template spectra, template SEDs and their corresponding gas-phase metallicity are compared with the observation through the following $\chi^2$-like likelihood function: 
\begin{equation}
\label{likelyhood:legac}
\begin{aligned}
\ln {L(\theta)}\propto &-
\frac{(Z_{\rm g, mod}-Z_{\rm g, obs})^2}{2\sigma_{Z}^2
}-\sum_{j=1}^M\frac{(F_{\rm mod,j}-F_{\rm obs,j})^2}{2F_{\rm err,j}^2}\\
&-\sum_{i=1}^N\frac{\left(f_{\rm mod,i}-f_{\rm obs,i}\right)^2}{2f_{\rm err,i}^2},
\end{aligned}
\end{equation}
In this formula $Z_{\rm g,mod}$ and $Z_{\rm g,obs}$ are model predictions and measured values of the current gas-phase metallicity, with $\sigma_{Z}$ being the estimated uncertainty. $F_{\rm mod, j}$, $F_{\rm obs, j}$ and $F_{\rm err,j}$ are the model flux, observed flux and error in the $j$-th photometric band, while $f_{\rm mod, i}$, $f_{\rm obs, i}$ and $f_{\rm err,i}$ are the model spectrum flux, the observed spectrum flux and flux error at the $i$-th wavelength point, with $M$/$N$ being the total number of photometric bands/wavelength points available.

By analysing the RMS variation of the stacked spectra, we found a median S/N=20 per pixel, so we use  $f_{\rm err,i}$=0.05$\times f_{\rm obs,i}$ for the entire wavelength range throughout the fitting. We found no major changes in the results when adopting a more conservative error choice, down to S/N=10 per pixel, based on the lower S/N spectral regions. Similarly, We adopt a typical photometric error of  $F_{\rm err,j}=0.02\times F_{\rm obs, j}$ \citep{Cappellari2022,Zhou2024legac} for the photometry bands. 

\begin{figure}
    \centering
    \includegraphics[width=0.5\textwidth]{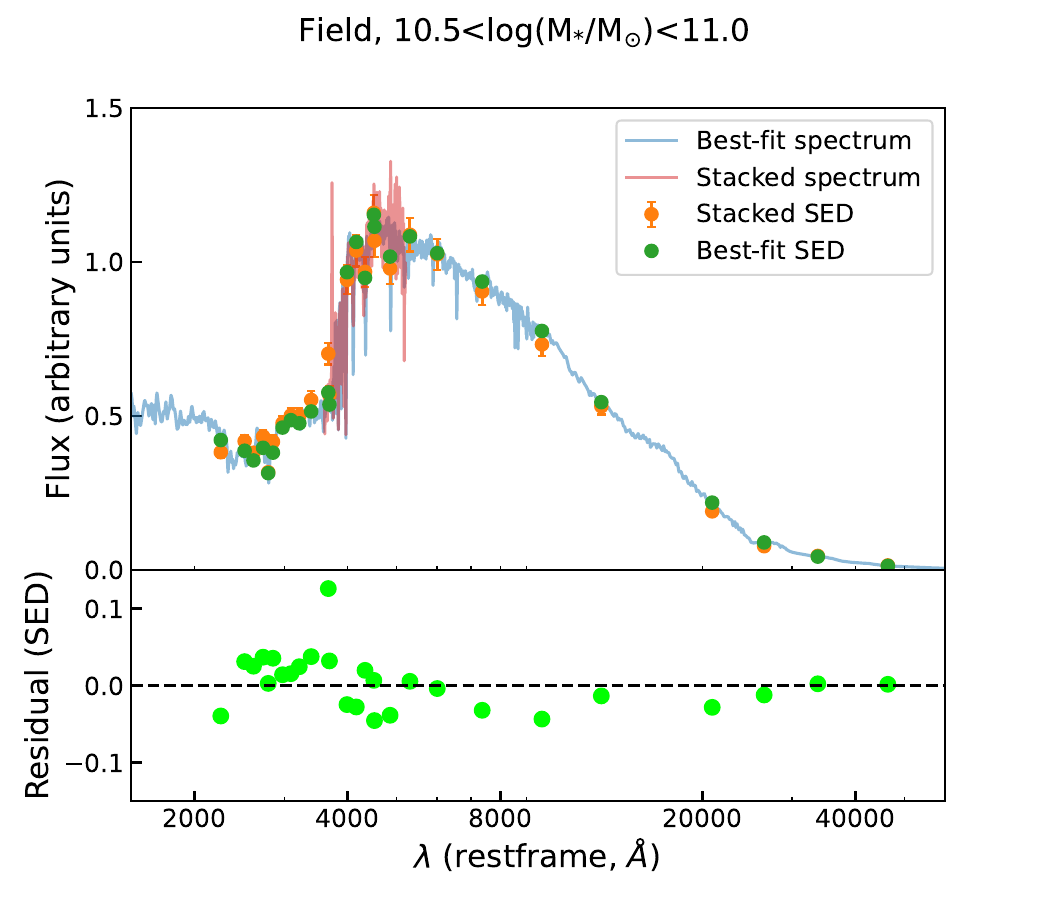}\\
    \includegraphics[width=0.5\textwidth]{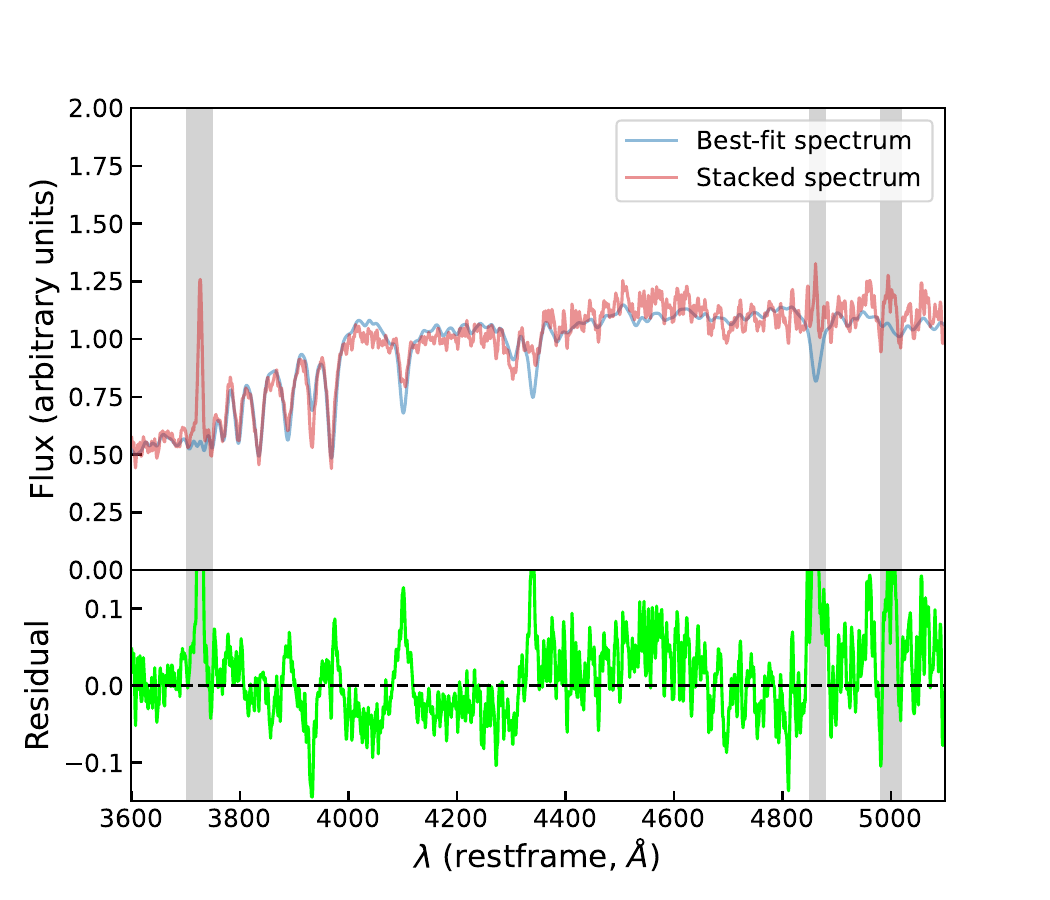}\\
     \caption{Example of this work's stellar population fitting process. The top panel shows our best-fit model spectrum (blue) and SED (green) compared to the stacked spectrum (green) and SED (orange), while the bottom panel zooms into the optical range. Residuals are shown for the SED in the top panel and for the spectrum in the bottom panel.
     Spectral regions around the emission lines are masked during the fitting, as indicated by the shaded regions in the bottom panel.}
     \label{fig:example_BIGS}
\end{figure}

The posterior probability distributions of the main physical parameters is calculated using the {\tt MULTINEST} sampler \citep{Feroz2009, Feroz2013} and its \textsc{Python} interface \citep{Buchner2014}. An example of the fitting output is shown in Fig.\ref{fig:example_BIGS}.
The best-fit SED closely follows the observed SED, while the best-fit spectrum appears to provide a worse match to the observed stacked spectrum than the pPXF fit shown in Fig.\ref{fig:example}. 
Several factors contribute to this difference. To begin with,  pPXF adopts multiple polynomials to adjust for the large-scale flux variations in the spectra, which ensures a perfect match between the continuum shape of the model and that of the observed spectrum. 
In addition, pPXF allows for arbitrary combinations of SSPs to fit the observations. Such flexibility is optimal for continuum subtraction purposes, 
yet it could lead to un-physical solutions 
involving combinations of age and metallicity values unlikely to occur in the real universe -- as discussed in \cite{WangZixian2023}, pPXF tends to overestimate the star formation rate (SFR) when the mean age of the stellar populations is between 2 and 4 Gyrs and between 10-14 Gyrs. 
In contrast, our model fitting aims to obtain physically consistent SFH and ChEH that follow the current galaxy evolution scenario, even if this implies losing the flexibility needed to match the details of the observed spectrum. Although our models are designed to capture the main physical processes of star formation, gas inflow, and outflow, we cannot exclude that some discrepancies may still arise due to the assumed priors and to physical processes that are not fully considered.
In addition, we model the stacked SED and spectrum simultaneously in our fitting. The spectrum data covers only the rest-frame blue end of the optical range, so it is more sensitive to the younger stellar population. In contrast, the wide wavelength coverage of the SED data allows the tracing of all the stellar components across all ages. 
Therefore, it is not surprising that including the SED in the fitting process can affect the fitting quality of the spectral optical range (but see \citet{Zhou2024legac} for a discussion of the main advantages of such a strategy).

As a quick sanity check, we derive the mass-to-light ratio in the r-band from both pPXF fitting and our analysis. The results are consistent within 0.18 dex, suggesting that our semi-analytic spectral fitting approach produces stellar populations consistent with those obtained from non-parametric spectral fitting methods.

\begin{table}
	\centering
	\caption{Priors of model parameters}
	\label{tab:paras}
	\begin{tabular}{lccr}
		\hline
		Parameter & Description & Prior range\\
		\hline
		$\tau$ & Gas infall timescale & $[0.0, 10.0]$Gyr\\
		$t_{0}$ & Start time of gas infall & $[0.0, 7.0]$Gyr\\
		$\lambda$ & The wind parameter & $[0.0, 10.0]$\\
		$E(B-V)$&  Dust attenuation parameter & $[0.0, 0.5]$\\
		\hline
	\end{tabular}
\end{table}

\section{Result and discussion}
\label{sec:results}

This section will present the physical information extracted from the stacked spectra and SEDs of the SFGs in the Wall Volume sample and their dependence on galaxy stellar mass and environment.  We first focus on results obtained on the instantaneous gas-phase properties from the nebular emission lines. 
We then show results on the past evolution of these galaxies as obtained from general empirical evidence and from our detailed spectral and SED model-fitting process. 
Finally, we discuss on the possible origins of the environmental dependence shown by our results.  

\subsection{Gas phase metallicity as a function of mass in different environments}
\label{ssec:result_gas}

We estimated gas-phase metallicities of galaxies using emission lines from the stacked spectra, in different bins of stellar masses and environments, following the approach described in Section \ref{subsec:gas_metal_measure}. 

\begin{figure}
    \centering
    \includegraphics[width=0.46\textwidth]{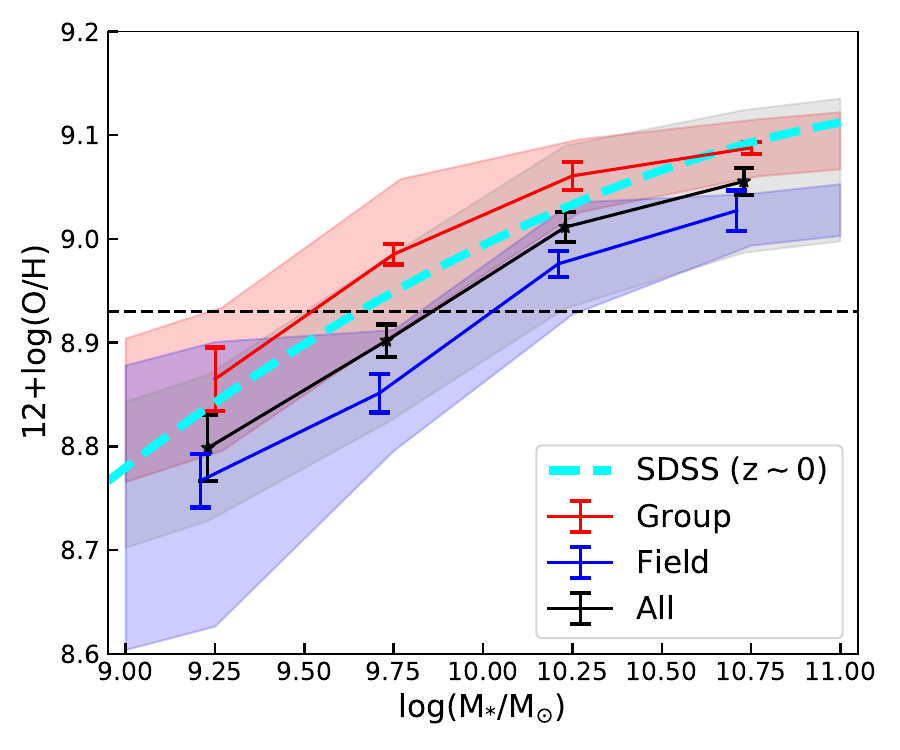}\\

     \caption{The gas-phase metallicities as a function of galaxy stellar mass and environment as obtained from the stacked spectra. Results from all galaxies in our sample are shown in black, while results from galaxies in groups are shown in red and field galaxies in blue. The error bar indicates the measured error obtained by mock repeat measurements, while the shaded region indicates the 1 $\sigma$ scatter of each sample estimated from bootstrapping. The cyan dash line indicates the MZR of local z$\sim$ 0 SDSS galaxies obtained by \cite{Tremonti2004}, while the horizontal dash line marks the solar oxygen abundance.}
     \label{fig:gasmetal}
\end{figure}

The results are shown in Fig. \ref{fig:gasmetal}, where different lines are the mass-metallicity relation of galaxies derived 
for different subsamples of the galaxy's stellar mass and environments. 
The error bars shown in Fig. \ref{fig:gasmetal} are the formal error bars, obtained taking into account only the noise of each stacked spectrum. 
To obtain these error bars, we first get an estimate of the typical noise value $N$ for each stacked spectrum by averaging the RMS variations in the spectral window near 4000{\AA} from the residuals obtained from the pPXF fits (see Fig.\ref{fig:example}).
Each error bar is then estimated from the standard deviation of repeated measurements of gas-phase metallicities on 100 spectra obtained by perturbing the best-fit model of each stacked spectrum with a noise contribution randomly generated from Gaussians of width equal to $N$. The shaded regions around each line in Fig. \ref{fig:gasmetal}, on the contrary, indicate the intrinsic scatter of the gas-phase metallicities estimates across the galaxies that inhabit the same bin of mass and environment. This scatter is estimated using a bootstrapping approach: for each stellar mass and environment bin containing $N_{gal}$ galaxies (as listed in table\ref{tab:bins}), we generate a bootstrap sample by randomly selecting $N_{gal}$ galaxies from it with replacement, i.e. allowing for the possibility of repeated galaxies in the bootstrapped samples. We generate 100 such bootstrapped samples for each bin and perform the same analysis on each of the bin. The shaded regions show the scatter of the results obtained this way. As expected, shaded regions are always bigger than the formal error bars, which take into account only the noise of the stacked spectra.

\begin{figure}
    \centering
    \includegraphics[width=0.45\textwidth]{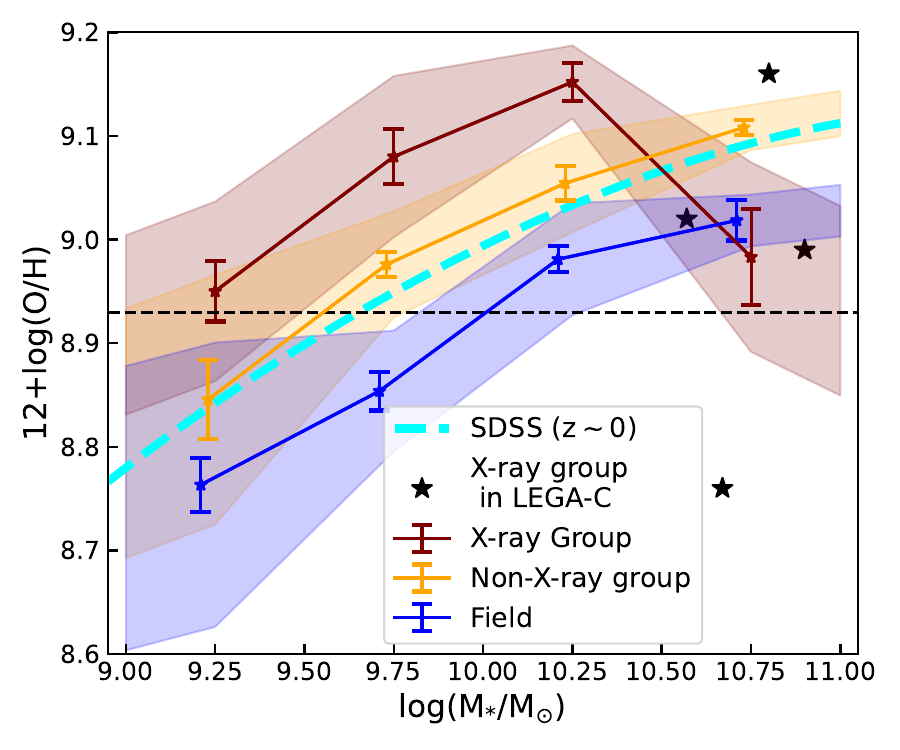}\\

     \caption{The gas-phase metallicities of our sample galaxies as a function of stellar mass obtained from the stacked spectra. Results for galaxies in X-ray groups are shown in dark red, galaxies in non-Xray groups are shown in orange and field galaxies in blue. The error bar indicates the measured error obtained by mock repeat measurements, while the shaded region indicates the 1 $\sigma$ scatter of each sample estimated from bootstrapping approaches. Black stars indicate individual measurements for X-ray group galaxies in the most massive bin available from high-quality LEGA-C observation. The cyan dash line indicates the MZR of local z$\sim$ 0 SDSS galaxies obtained by \cite{Tremonti2004}, while the horizontal dash line marks the solar oxygen abundance.}
     \label{fig:gasmetal_xray}
\end{figure}

The black line in Fig. \ref{fig:gasmetal} shows the mass-metallicity relation (MZR) obtained from the stacked spectra of all galaxies, binned in stellar mass, in our sample. Thanks to the good mass coverage of the Wall Volume sample, our derived MZR spans a wide dynamical range from 10$^{9}$M$_{\odot}$ to 10$^{11}$M$_{\odot}$. The MZR we obtained follows the well-known trend where more massive galaxies have higher gas-phase metallicities.  
As a reference, the cyan dashed line in Fig. \ref{fig:gasmetal} shows the present-day MZR of z$\sim$0 SDSS galaxies obtained by \cite{Tremonti2004}. Note that the \cite{Tremonti2004} results are based on a different gas-phase metallicity calibration, but we have converted their results to the KK04 calibration using the formula provided by \cite{Kewley2008}. Compared to this present-day MZR, our mean MZR, for galaxies at z$\sim0.73$, falls on average 0.04 dex below. This result is well consistent with what has been found by other investigations of galaxies at z$\sim$0.7 such as works based on the DEEP2 survey \citep{Zahid2011} and the LEGA-C survey \citep{Lewis2024}. 

In addition to this global trend, in this work, we have detailed information on the environment of our sample galaxies, which allows us to explore the potential influence of the environment on the evolution of galaxies. To this end, in Fig. \ref{fig:gasmetal}, we plot the MZR obtained from stacked spectra of galaxies in groups (magenta) and in the field (blue), respectively. It is immediately visible that galaxies in groups are on average $\sim$ 0.09 dex higher in gas-phase metallicities compared to galaxies in the field in all the four stellar mass bins adopted in this work. Such an environmental dependence is well in line with local investigations. For example, from a sample of galaxies selected from SDSS, \cite{Cooper2008} find that galaxies in high-density regions tend to have higher gas-phase metallicities. \cite{Pasquali2012} further reveal that, for local galaxies, being a satellite in a group can induce a 0.06 dex increase in its gas phase metallicity compared to that of the central galaxies of similar stellar mass.
Our galaxies in groups reside in denser environments, and many of them are satellite galaxies. The consistency of results from our samples at intermediate redshift and local measurements suggests that at a given stellar mass, the environment's non-negligible impact on the galaxy's chemical evolution was already in place $\sim$ 7 Gyrs ago.

Beyond a simple division of galaxies into group and field galaxies, we can further separate our group's sample galaxies according to whether the group they belong to has or not an X-ray counterpart (see Section \ref{ssec:envir}). In Fig. \ref{fig:gasmetal_xray}, we plot the MZR for galaxies residing in groups with X-ray detection (X-ray groups), galaxies residing in groups without X-ray detection (non-X-ray groups) and field galaxies in dark red, orange and blue lines, respectively. Similar to Fig. \ref{fig:gasmetal}, the error bars in the plot show the formal measured errors while the shaded regions represent the 1 $\sigma$ scatter in each bin as obtained from the bootstrapping procedure. Galaxies in X-ray groups tend to have even higher gas-phase metallicities than those in non-X-ray groups. Our results thus could indicate that galaxies experienced more significant environmental effects during their evolution when located in X-ray emitting groups, which are the richer and presumably more massive groups. 

It is noticeable that galaxies in X-ray groups and in the highest mass bin ($10^{11.0}>M_{*}/{\rm M}_{\odot}>10^{10.5}$) display gas-phase metallicity values comparable to or even slightly smaller than those of field galaxies. To better understand such a result, we looked in detail at the distribution of the derived gas-phase metallicities from the 100 bootstrapped samples. Such distribution displays a long tail towards the lower gas-phase metallicities values, extending down to more than 0.3 dex below the average value. Such a significant variation indicates the existence of a population of massive metal-poor galaxies in this mass bin for galaxies residing in X-ray-emitting groups. As a further test, we matched our X-ray group sample with galaxies from the LEGA-C survey \citep{vanderWel2016, Straatman2018}. The match enables us to identify four galaxies with LEGA-C spectra, with $10^{11.0}>M_{*}/{\rm M}_{\odot}>10^{10.5}$ and located in X-ray emitting groups. While the S/N of individual galaxies spectra in our sample are insufficient to obtain reliable measurements of the gas-phase metallicities, LEGA-C data spectra enabled us to measure gas-phase metallicities for each of these four galaxies. The results are shown as black stars in Fig. \ref{fig:gasmetal}. Although the number of galaxies is too small for statistical analysis, we again see a wide spread of gas-phase metallicity values extending to quite low values. The matching our sample galaxies in non-X-ray groups and in the field to the LEGA-C sample, although not shown in the figure, shows a smaller spread of the gas-phase metallicity values, but the number of matches is too small to draw a solid conclusion. The existence of such a population of massive low metallicity galaxies at redshift $z\sim0.7$ has already been noticed by many previous works \citep[e.g.][]{Zahid2011, Maier2015, Huang2019, Lewis2024, Zhou2024legac}, that, however, did not state any environmental dependency. Our results suggest that the denser group environment may have contributed to the formation of such galaxies. We will come back and discuss the possible origins of these galaxies in Section \ref{subsec:origin}.

In summary, the gas-phase metallicities obtained from emission lines agree well with the MZR obtained in most previous investigations at intermediate redshift. Thanks to the availability of environmental information, we further reveal a clear dependence of the MZR on the environment so that galaxies in groups have higher gas-phase metallicities than galaxies in the field. In addition, X-ray groups, $i.e.$, the more massive groups, have a more significant impact on the gas-phase metallicities of the member galaxies. An interesting exception to this trend is for the most massive stellar-mass bin in the X-ray emitting groups, where a population of massive low metallicity galaxies is observed. 
As the evolution of gas and stars are deeply entangled during a galaxy's life cycle, it is interesting to explore whether such environmental effects are also imprinted in the stellar population of galaxies. In what follows, we will turn to the stellar population properties of our sample galaxies and discuss the past evolution of these galaxies. Given the interesting physical insights provided by the X-ray group/non-X-ray group/field division, in what follows, we will stick to these three samples and try to find more clues on environmental effects on galaxy evolution.

\subsection{Stellar population properties as a function of mass in different environments}

\subsubsection{Empirical evidence}

\begin{figure}
    \centering
    \includegraphics[width=0.45\textwidth]{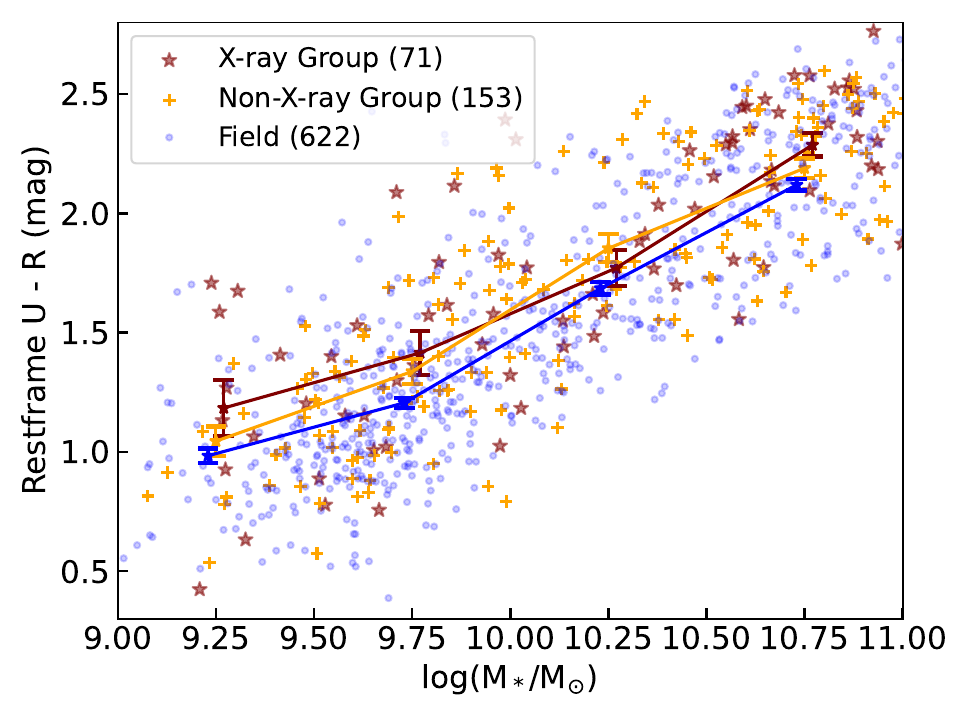}\\
    \includegraphics[width=0.45\textwidth]{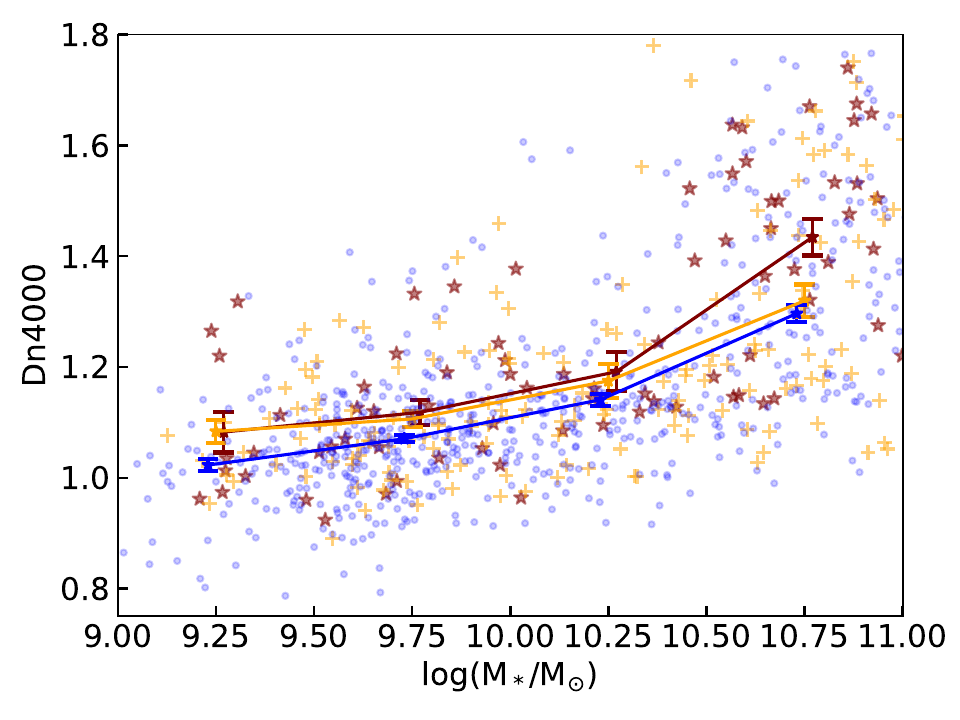}\\
    \includegraphics[width=0.45\textwidth]{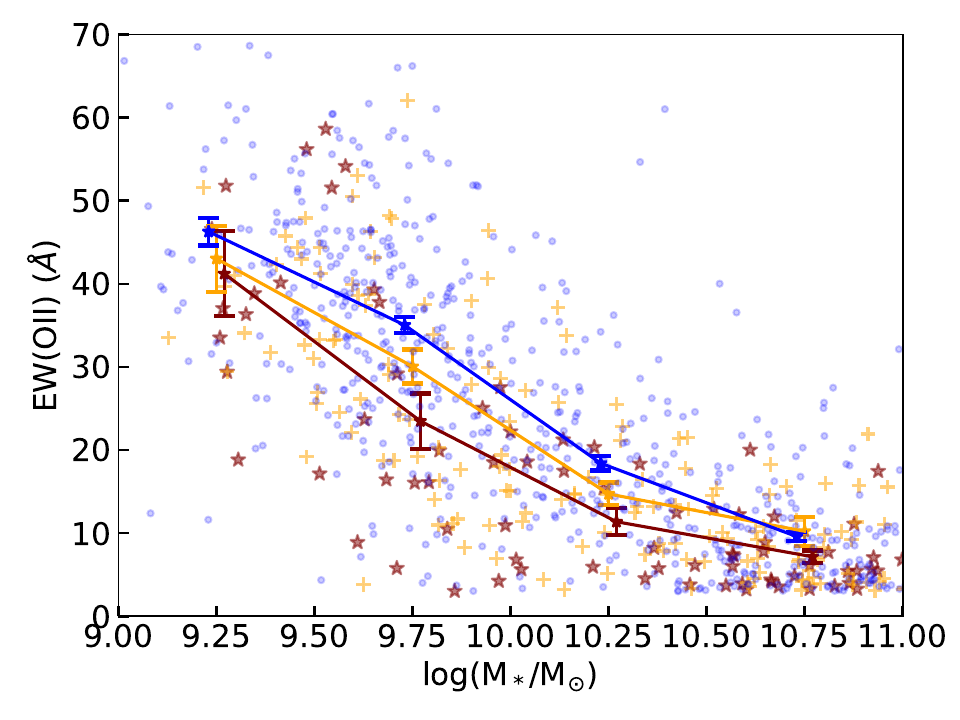}\\
     \caption{The rest-frame U-R colour (top), Dn4000 index (middle) and  \hbox{[O\,{\sc ii}]} equivalent width (bottom) of our sample galaxies as a function of stellar mass. Galaxies in X-ray groups are shown with dark red stars, while galaxies in non-X-ray groups are shown with orange crosses and field galaxies with blue dots. The legend on the top panel indicates the total number of galaxies for each category. Colour lines show the mean values in the stellar mass bins of the corresponding category of galaxies, with error bars indicating the 1$\sigma$ error of the mean value.}
     \label{fig:stp}
\end{figure}

We can now explore how the stellar population properties of our sample of SFGs depend on the environment. To achieve this, we first focus on some empirical evidence that allows us to investigate such a dependence in a model-independent way. 

The colour of a galaxy is a first order indicator of its stellar population content. Galaxies with young stellar populations are generally bluer, while older stellar populations are redder in colours \citep{Kauffmann2003}. To get a broad view of the stellar population properties of our SFG sample, we used their rest frame U-R colour from the Wall Volume catalogue (see section \ref{ssec:Ssel}).

The top panel of Fig.\ref{fig:stp} shows the restframe U-R colour as a function of stellar mass for our SFG sample. Galaxies from X-ray groups, non-X-ray groups, and field galaxies are shown with dark red stars, orange crosses, and blue dots. The average trend of galaxies in different environments is marked by a solid line, with error bars showing the uncertainties on the mean values. The plot shows that more massive galaxies in our sample are redder in their U-R colours, indicating an older stellar population.
In addition, we observe a secondary colour dependence on the environment. At a given stellar mass, galaxies in groups, especially X-ray groups, are slightly redder than field galaxies. 
This result suggests that group galaxies tend to have older stellar populations, which indicates that galaxies in different environments have non-negligible differences in their evolution paths.

We further confirm these findings by looking into other spectral features. In the second panel of Fig.\ref{fig:stp}, the y-axis is replaced by the Dn4000 index, a good indicator of the stellar age of the galaxy, such that galaxies with older stellar populations tend to have higher Dn4000 indices. Unsurprisingly, we see a trend similar to that of the top panel, that is group galaxies, on average, have older stellar populations than field galaxies. Finally, the bottom panel of Fig.\ref{fig:stp} shows the distribution of the equivalent width of the \hbox{[O\,{\sc ii}]} emission line, a good proxy for the specific star formation rate of a galaxy. At a given stellar mass, higher \hbox{[O\,{\sc ii}]} equivalent width indicates stronger recent star formation activity. We can thus infer from the plot that SFGs in groups, especially X-ray groups, have systematically weaker recent star formation activity, which is in line with their older stellar populations indicated by the top and middle panels.

This empirical evidence suggests that the environment affects not only the current gas-phase metallicity of the galaxies but also their past star formation history, as revealed by their stellar population properties. In what follows, we will use the results obtained from a detailed analysis of the spectra and SEDs of our sample galaxies to seek more evidence of environmental effects, as well as to understand the possible physical mechanisms causing them.

\begin{figure*}
    \centering
    \includegraphics[width=0.8\textwidth]{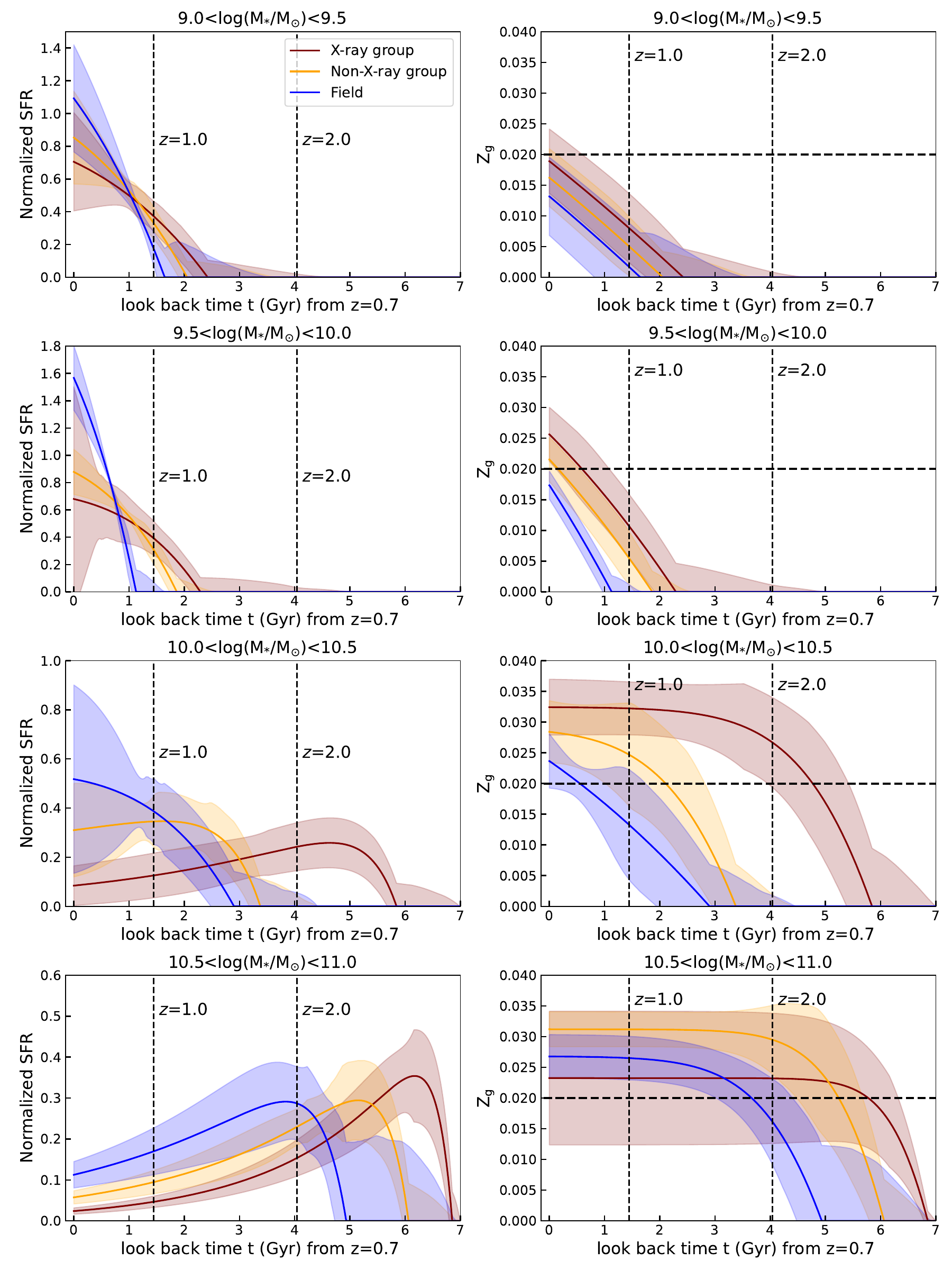}\\
     \caption{The SFHs (left) and ChEHs (right) of our sample galaxies, as obtained from best-fit models of the stacked spectra and SEDs. From top to bottom, different rows show results from different stellar mass bins, as indicated. Results from galaxies in X-ray groups are shown with dark red lines, while galaxies in non-X-ray groups are shown in orange and field galaxies in blue. Shaded regions are 1$\sigma$ scatters estimated from the bootstrapping process. Two vertical dashed lines indicate the position of redshift $z=1.0$ and $z=2.0$, respectively. The horizontal dashed line marks the solar metallicity used in this work.}
     \label{fig:result_sp}
\end{figure*}

\subsubsection{Results from spectral analysis}

We now apply our semi-analytic fitting approach to the spectra and SEDs of our SFG sample. 

In Fig.\ref{fig:result_sp}, we plot the best-fit SFHs and ChEHs obtained from the stacked spectra and SEDs for galaxies in different stellar mass bins and environments. In each panel, results obtained for X-ray groups/non-X-ray groups/field galaxies are plotted in dark red/orange/blue, respectively, while shaded regions around lines are 1 $\sigma$ variation obtained from the bootstrapping analysis.  The x-axis is set to show the look-back time from $z\sim 0.7$, with the two vertical dashed lines indicating the position of redshift $z=1.0$ and $z=2.0$, respectively. 

The results for the least massive galaxies ($10^{9.0}<M_{*}/{\rm M}_{\odot}<10^{9.5}$) are shown in the top panels of Fig.\ref{fig:result_sp}. Their SFHs are shown in the top-left panel, and we observe a clear trend that galaxies in groups form systematically earlier than field galaxies. This result is consistent with 
the empirical evidence discussed in the previous section, where galaxies in groups have higher Dn4000 values and lower \hbox{[O\,{\sc ii}]} equivalent widths. A similar evolution is also found in their chemical compositions. As shown in the top right panel, galaxies in groups, especially X-ray emitting, massive groups, start to be enriched in metals earlier, which leads to the higher gas-phase metallicity in group galaxies as observed in Fig.\ref{fig:gasmetal}. In addition, we can notice that, in these galaxies, their star formation activity and chemical enrichment process are still increasing at the time of observations, $i.e.$ at $z\sim 0.7$. Similar evolutionary trends, as well as the dependence on the environment, are also found in the second stellar mass bin ($10^{9.5}<M_{*}/$M$_{\odot}<10^{10.0}$).

This evolution picture starts to change when we come to more massive SFGs. In the third stellar mass bin ($10^{10.0}<M_{*}/{\rm M}_{\odot}<10^{10.5}$) as shown in the third row of Fig.\ref{fig:result_sp}, we found that galaxies in X-ray groups (dark red), which formed earlier, are already undergoing star formation quenching at $z\sim 0.7$. The right panel shows that their gas-phase metallicity reaches an equilibrium state and forms a plateau already 4 Gyrs before $z\sim 0.7$. In contrast, field galaxies in this stellar mass bin (blue) follow an evolution path similar to that found in the less massive bins, while non-X-ray group galaxies are in a transition state between the two categories. It is well known 
that massive galaxies have different SFHs compared to low mass ones \citep[e.g.][]{Panter2003, Kauffmann2003, Heavens2004, Panter2007, Fontanot2009, Peng2010, Muzzin2013, Zhou2022}. Our results clearly show that such a transition correlates with the environment. At a given stellar mass, galaxies in denser environments evolve faster and display an earlier decline in their star formation activities than field galaxies. Their chemical enrichment processes also reflect this variation of speed in star formation activity. While galaxies in X-ray groups reach the plateau of equilibrium in their gas-phase metallicities, galaxies in fields at  $z\sim 0.7$ are still becoming enriched in their metal content.

Finally, the last row shows the most massive galaxies of our sample. In this mass bin, galaxies in all types of environments reached the peak of their star-formation activity more than 4Gyrs before $z\sim 0.7$ and, since then, display a decreasing star formation rate. The impact of the environment is still clearly visible, as galaxies in X-ray groups reached the peak of their 
SFH  
systematically earlier and have shorter star formation timescales. Regarding the chemical evolution, galaxies in all the environments grow their metallicity in the first 2-3 Gyrs of their evolution, reaching a value similar to the observed present-day gas-phase metallicity, as measured by the emission lines ratios in Section \ref{ssec:result_gas}. 
They then get into the equilibrium state so that their gas phase metallicities remain unchanged since $z\sim1-2$ depending on their environment (earlier for denser environments). It is intriguing to see that, despite reaching a similar mass through earlier and faster growth, galaxies in X-ray groups maintain relatively lower metallicity values
since $z\sim 2$, which indicates that another physical mechanism rather than simply the star formation time and time-scale  
may play a role in shaping their chemical evolution. 

In summary, the SFHs and ChEHs of our sample galaxies derived from best-fit models to the stacked spectra well align with empirical evidence obtained from emission line measurements and spectral indices,   suggesting that our models have succeeded in capturing the main physics governing the evolution of these galaxies. We now move to discuss some of the potential physical interpretations of our results.

\subsection{Origin of the environment dependence}
\label{subsec:origin}
Our semi-analytic spectral fitting approach enables us to retrieve the SFHs and ChEHs of our sample of SFG in different environments, based on a simple model where the gas inflow and outflow histories largely determine how a galaxy has evolved. 
In this section, we go back to the basic parameters of our model 
to discuss a possible physical interpretation of our results. 

\begin{figure*}
    \centering
    \includegraphics[width=1.0\textwidth]{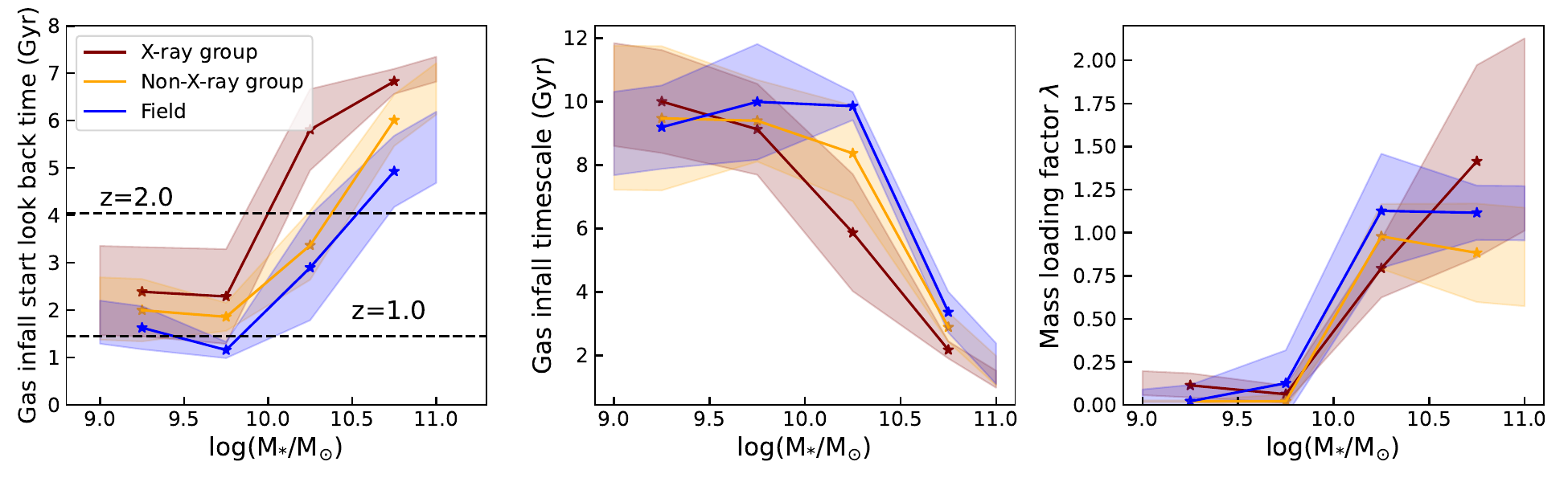}\\
     \caption{The start time of the gas infall (left), gas infall time scale (middle) and outflow strength as indicated by the wind parameter factor (right) of our sample galaxies as functions of their stellar mass, obtained from best-fit models of the stacked spectra and SEDs. Results from galaxies in X-ray groups are shown with dark red lines, while galaxies in non-X-ray groups are shown in orange and field galaxies in blue. Shaded regions are 1$\sigma$ scatters estimated from bootstrapping. }
     \label{fig:result_para}
\end{figure*}

In Fig.\ref{fig:result_para}, we plot the derived gas infall starting time (left), gas infall timescale (middle) and outflow strength or wind parameter factor (right) as a function of the stellar mass bin for galaxies in different environments.
The gas infall starting time indicates the beginning of the galaxy's star formation activities, which in our simple model can be regarded as the starting point of galaxy formation. The Wall galaxies are observed at around redshift $z\sim$0.7, when the age of the Universe was roughly $\sim$7~Gyrs, and the left panel of Fig.\ref{fig:result_para} shows the starting time of the gas infall as look-back time from redshift $z\sim$0.7, so that higher values indicate earlier gas infall. For references, we also indicate the look-back time corresponding to $z=$1.0 and $z=2.0$ with horizontal dashed lines. The timescale of such gas infall, as shown by the central panel of Fig.\ref{fig:result_para}, characterises the duration of the gas-feeding process: a long gas infall timescale indicates continuous gas supply to the galaxy, which enables long-term star formation activity. In addition, the long-lasting infall of pristine gas can dilute the ISM, slowing down the chemical enrichment process in a galaxy. On the other hand, the outflow tends to remove chemical-enriched gas from the galaxy. Strong outflow, as parametrised by higher $\lambda$ wind parameter values, would quickly blow away the gas reservoir in a galaxy, shutting down the star formation process and suppressing the chemical enrichment process.

The first panel of Fig.\ref{fig:result_para} shows a consistent trend: galaxies in denser environments have a higher gas infall starting look-back time, i.e. these galaxies began their star formation activity at earlier, more remote, epochs. This is well in agreement with a hierarchical structure formation scenario, where the richer groups observed at $z\sim0.7$ correspond statistically to the first collapsed structures in the Universe. In parallel, as galaxy mass increases, the gas infall timescale progressively decreases, and the trend with the environment is for shorter time scales in denser environments.
For the lowest mass galaxies, the value of  gas infall timescale appear to reach the upper boundary of our prior settings (10 Gyr), irrespective of their environment. Similarly, their outflow strength is close to zero, irrespective of their environment. 

This lack of environmental dependence for the lower mass bins galaxies suggests that the hierarchical build-up scenario is sufficient to explain their evolution: a possible interpretation is that  
the galaxies in these lowest mass bins still exhibit increasing SFRs at the time of observation ($z\sim 0.7$), indicating they retain sufficient gas and have not been heavily affected by interactions with the group environment. For these galaxies, our model cannot define the timescale for the (yet to happen) decline in gas inflow, as the system is still actively experiencing growth in its star formation activity.

On the contrary, for the two most massive stellar mass bins, i.e. $M_{*}/{\rm M}_{\odot}>10^{10.0}$, the star formation activity of galaxies and the gas infall is already in its declining phase at $z\sim 0.7$ while the outflow strength is substantial. As far as the environment dependence is concerned, massive galaxies in lower density regions display a longer gas infall time scale, while galaxies in denser regions display a shorter infall time scale and thus an earlier cessation of star formation activity. The picture for the outflow strength is more complex, as it shows a marked difference in the dependence on environment between the two most massive stellar mass bins.

In contrast to the low mass galaxies, such a variation of gas infall time scale and outflow 
in high mass galaxies is difficult to explain solely through a simple hierarchical build-up scenario. Additional environmental effects have to be considered to account for these results.

The early cessation of star formation activity for galaxies in groups has long been noticed in the literature, and the lack of infalling gas is proposed as a possible driver for such a trend. Using large-scale surveys, several investigations reveal that galaxies in groups and clusters \citep[e.g.][]{Pasquali2010, Peng2012, Wetzel2012, Wetzel2013} and in high-density regions \citep[e.g.][]{Kauffmann2003, Baldry2006} tend to be more quenched in their star formation activity compared to isolated galaxies. In current galaxy formation scenarios, satellites falling into a more massive dark matter halo may experience the removal of their hot gaseous halo, i.e. the so-called `strangulation' process \citep{Balogh2000, Pasquali2010}, so that the galaxies get quenched with the consumption of their cold gas reservoir. Alternatively, direct ram-pressure stripping of their cold star-forming gas can also result in an early end to their gas supply, with an even shorter timescale \citep{Gunn1972, Abadi1999}. Compared to these studies for mass complete samples in the local Universe, our analysis focuses on star-forming galaxies observed at $z\sim0.7$. However, our results here provide similar evidence indicating that the environmental effect was already visible in star-forming galaxies at least 7~Gyrs ago. 
 
As far as gas metallicity is concerned, in our model, the longer-lasting infall of pristine gas in field galaxies should dilute their ISM so that their gas-phase metallicities are kept at a lower value compared to group galaxies. 
This effect is well visible for galaxies in the mass range  $10^{10.0}$<$M_{*}/{\rm M}_{\odot}<10^{10.5}$, where the expected trend in gas metallicity is visible in Fig. 11. 

Moving to the most massive stellar-mass bin ($10^{10.5}<M_{*}/{\rm M}_{\odot}<10^{11.0}$) and to galaxies in X-ray groups, the current gas-phase metallicity of these galaxies, see Fig.\ref{fig:gasmetal_xray}, instead of being higher, drops by around 0.2 dex compared to the less massive bin. This drop makes their gas-phase metallicities comparable or even slightly lower than field galaxies in the same stellar mass bin.

The existence of massive metal-poor galaxies has been documented in numerous studies, in both gas-phase \citep[e.g.][]{Zahid2011, Maier2015, Huang2019} and in stellar phase \citep[e.g.][]{Gallazzi2014, Beverage2021}, and is also predicted by semi-analytical models \citep{DeLucia2012}. Our observational results confirm these previous findings and further reveal that this metal deficiency is preferentially observed in galaxies in richer groups.

A number of explanations could account for the metal deficiency in these galaxies, including but not limited to the additional infall of pristine gas, removal of chemically enriched gas, and recent galaxy mergers \citep{DeLucia2012}. Although the limited data quality and sample size prevent us from fully exploring all these possibilities, we can make some rough inferences based on our current results. For instance, our model does not explicitly include multiple epochs of gas infall, so we cannot directly constrain the recent infall of pristine gas. However, if the metal deficiency in these massive galaxies were due to recent pristine gas infall, we would expect to observe a co-existence of old and young stellar populations, which would effectively lengthen the star formation timescales. However, this is not observed in our results. Instead, we find that X-ray group galaxies formed earlier and quenched their star formation more rapidly. The short gas infall timescale suggests that pristine gas infall is unlikely to account for the low gas-phase metallicities in these galaxies.  To investigate whether recent mergers could be responsible for this effect, we visually inspected the HST images of these galaxies and confirmed that none contain very recent merger remnants. In fact, galaxies showing significant mergers or interactions with close companions were excluded during our sample selection process (see Section \ref{sec:data}).

Alternatively, the efficient removal of chemically enriched gas seems to be a plausible explanation for this phenomenon. In their study of passive galaxies from LEGA-C, \cite{Beverage2021} suggest that the low-metallicity galaxies they identify using a conventional spectral fitting approach could be explained by the effective removal of gas from these systems. In our sample, the right panel of Fig.\ref{fig:result_para} shows that the mass-loading factor $\lambda$, which characterises the strength of gas removal relative to star formation activity, spans a wide range in X-ray group galaxies, with a mean value higher than that in smaller groups or the field. This result suggests that some of the galaxies in the richest groups may have experienced significant gas removal during their evolution, leading to their low current gas-phase metallicities. The similar finding from fairly different datasets strengthens the case for gas removal as a key process in the origin of these massive, metal-poor galaxies at $z\sim0.7$ \citep{Beverage2021}. Our model itself does not provide further insight into the physical processes responsible for gas removal. However, based on our findings, we can briefly discuss the most likely mechanisms responsible for such a phenomenon.

One possible scenario is the ram-pressure stripping -- when the galaxy group's potential well is deep enough, direct ram-pressure stripping can remove the gas reservoir when a satellite galaxy falls into the group \citep{Gunn1972, Abadi1999}. Such a direct stripping will lead to a fast quenching of the galaxies' star formation activity and make them relatively metal-poor due to the loss of metal-enriched gas. 

Note that we impose an upper mass cut of M$_*$<10$^{11}$M$_{\odot}$ for the most massive galaxies. In this case, all of the galaxies in the X-ray groups within this mass bin are satellite galaxies, as the more massive central galaxies have been excluded. It is, therefore, reasonable to expect that galaxies in more massive X-ray-emitting groups are more likely to experience such processes. Still, it may be hard to understand why such an effect is seen preferentially in massive galaxies.

Strong AGN feedback can also effectively remove gas from galaxies, inducing a similar effect on their evolution. Some simulation works have revealed such a possibility. \cite{Torrey2019} show that IllustrisTNG simulations predict low metallicity for galaxies more massive than $10^{10.6}{\rm M}_{\odot}$ at $z\sim1$, which is due to the strong AGN feedback recipe used. \cite{Zhou2024legac} also found in a sample of massive metal-poor galaxies observed by the LEGA-C survey more residual AGN activity. However, although the stellar mass of the most massive bin of our findings meets the average critical mass ($\sim 10^{10.5}{\rm M_{\odot}}$) where AGN feedback is thought to be important \citep{King2015}, it is not easy to explain why such AGN activity is more prominent in X-ray emitting group galaxies.

Due to the limited size and data quality of the sample used in this work, not to mention the simplistic models used, we cannot shed further light on the exact origin of such strong outflow or discuss alternative mechanisms further.  In the future, large surveys with high-quality spectra of individual sources at this intermediate redshift range, such as WEAVE-Steps \citep{Iovino2023}, will provide a more detailed view and may help to better clarify the physical origin of these galaxies.

\section{Summary} 
\label{sec:summary}
In this work, we investigated the environmental dependence of the formation and evolution of 846 star-forming galaxies at intermediate redshifts ($z\sim0.73$) selected from the Wall Volume dataset. We obtained stacked spectra and SEDs within small stellar mass bins for galaxies inhabiting different environments. We measured the emission line fluxes from the stacked spectra, which are converted to gas-phase metallicities. Combining constraints from the gas-phase metallicities, stacked spectra and SEDs, we derived models that best characterise the star formation and chemical evolution of galaxies through a semi-analytic spectral fitting process. This approach provides us with self-consistent SFHs and ChEHs, showing clearly how our sample galaxies formed and evolved and how their evolution has been affected by the environment they inhabited. In addition, the fitting also determines the physical parameters of gas accretion and outflow that generate the model, which helps to interpret the physical processes that drive such evolution and its environmental dependence. 

The main results of this work are as follows:

\begin{itemize}

\item We find that the gas-phase metallicity derived from the stacked spectra of star-forming galaxies within the Wall Volume positively correlates, on average, with the stellar mass of the galaxies, forming an MZR that is slightly ($\sim$0.05 dex) below the present-day MZR obtained from local SDSS galaxies. This finding agrees well with many of the investigations at $z\sim$0.7 based on other datasets.

\item We reveal that the MZR at $z\sim$0.7 has a clear environmental dependence. At a given stellar mass, galaxies inhabiting groups tend to have higher gas-phase metallicities than field galaxies. Galaxies in X-ray-detected groups are even more metal-rich than those in smaller, non-X-ray-detected groups, indicating stronger environmental effects from more massive, richer groups. However, the most massive galaxies in X-ray-detected groups do not follow this trend, displaying a broader spread in their gas-phase metallicities.

\item We also observe empirical evidence that the stellar population properties of galaxies depend on the environment, including the restframe U-R colour, Dn4000 index and equivalent width of the \hbox{[O\,{\sc ii}]} emission line. This evidence consistently indicates that galaxies in groups, especially in X-ray-detected massive groups, tend to have older ages and weaker current star formation activities. 

\item The SFHs derived from the semi-analytical fitting of the stacked spectra and SEDs confirm the trends revealed by the empirical SED/spectral evidence. Galaxies in groups, especially X-ray-emitting massive groups, are found to start their star formation earlier than field galaxies and to have systematically shorter star formation timescales. This is a consistent trend observed across all galaxies masses.

\item In the three lower mass bins ($10^{9.0}<M_{*}/{\rm M}_{\odot}<10^{10.5}$), an earlier formation time and  a faster star formation timescale for galaxies located in denser environments allow them to accumulate metals earlier and more efficiently. 
The shorter gas infall timescale in these galaxies also mitigates the dilution of the ISM by pristine gas infall,
leading to higher chemical enrichment  compared to field galaxies. 
This shorter gas infall timescale indicates that the dense group environment may have stopped the cool gas from falling into the member galaxies. However, the earlier formation time in denser environments could also imply that star formation occurred in different internal physical conditions, leading to a more efficient star formation activity (and thus an earlier exhaustion of gas supply) even in the absence of an active role of the environment.

\item The situation is different in the highest mass bin and in the densest X-ray group environment. In this case, these parameters alone do not fully determine the galaxy chemical composition. In fact, galaxies in the highest mass bin and located massive in X-ray groups, despite their shorter gas infall time and shorter gas infall timescales, display a significantly lower gas metallicity and a markedly higher value of the gas outflow strength parameter measured from the best-fit models. We speculate that either direct ram-pressure stripping or strong AGN activity in these galaxies may have expelled chemically enriched gas, keeping them relatively metal-poor. However, it is not easy to explain why such AGN activity is more prominent in X-ray emitting group galaxies, so this result could hint at a possible role of the environment in these galaxies evolution, although without providing definitive proof.

\end{itemize}

\begin{acknowledgements}
This project was made possible by INAF funding through the PRIN 2011 program, the ASI, and the Italian Ministry grant `Premiale MITIC 2017'.  
The authors wish to acknowledge the generous support of ESO staff during service observations.
S.Z., A.I., M.L., M.S., S.B., M.B., O.C., L.P., D.V., E.Z. and F.D. acknowledge financial support from INAF Large Grant 2022, FFO 1.05.01.86.16. 
\end{acknowledgements}

\bibliographystyle{aa}
\bibliography{szhou}

\begin{appendix} 
\section{Measurements and uncertainty estimates of EW(OII)}
\label{app:measure}
In this appendix we present in detail how the EW(OII) values of the Wall galaxies are obtained and discuss the uncertainty level of the measurement. To begin with, each individual spectrum from a Wall galaxy is corrected to the rest-frame using the spectroscopic redshift measurement provided in \cite{Iovino2016}. We then apply a pPXF fit to the rest-frame spectrum to obtain the continuum shape. A Gaussian profile is fitted to the continuum-subtracted spectrum around restframe 3727 {\AA} to model the \hbox{[O\,{\sc ii}]}$\lambda$3727 emission line profile. Details and examples to the spectra fitting and line modelling procedure can be found in Section \ref{subsec:gas_metal_measure}. With both the continuum flux level ($F_c$) and flux in the emission line region ($F$) modeled, we then measure $EW(OII)$ in the wavelength range 3716.3 – 3738.3 {\AA} \citep{Belfiore2019} as \begin{equation}
    EW (OII) = -\int^{3738.3}_{3716.3} (1-\frac{F}{F_c})d\lambda.
\end{equation}
Note that for convenience in this work we define emission lines ($F>F_c$) to have positive equivalent width values.

\begin{figure}
    \centering

    \includegraphics[width=0.5\textwidth]{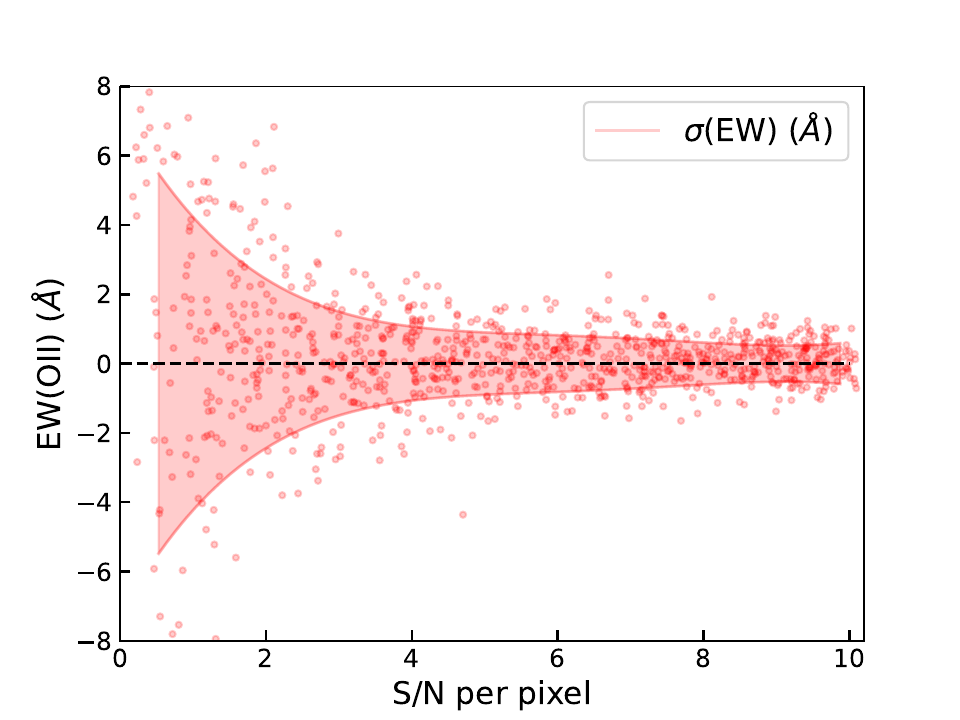}\\
     \caption{The measured $EW(OII)$ from a mock sample with no emission lines as a function of S/N. Each red dot is a measurement of a mock sample, with a black dash line indicating $EW(OII)=0$. The red shaded region shows the 1 $\sigma$ scattering level of the measurement.}
     \label{fig:mock_EW}
\end{figure}

The uncertainty of such a measuring process can not be obtained in a straightforward way. To assess the quality of our measurement, we perform a mock analysis as follows. To begin with, we obtain 1000 best-fit continuum spectra of the Wall galaxies from the pPXF fitting mentioned above to construct a mock galaxy sample. Such a mock sample contains spectral representatives of the continuum shape of our sample galaxies, with no emission lines presented. We then add random noise into the mock spectra so that their S/N varies from 0.1 to 10 per pixel. We repeat the measurements of EW(OII) on this mock sample and the results are shown in Fig.\ref{fig:mock_EW}. 

As shown in Fig.\ref{fig:mock_EW}, the EW (OII) measurements scatter around 0 due to fluctuations in the noise. We then calculate the standard deviation of such fluctuation $\sigma$(EW), and show it as the shaded region in the plot. This fluctuation can then be regarded as a representative of the uncertainty level of our measurement. From Fig.\ref{fig:mock_EW}, we see that the uncertainty is proportional to 1/(S/N), which is in line with the theoretical estimations of the uncertainty in simpler equivalent width measurements discussed in \cite{Vollmann2006}. In addition, as one would naturally expect, equation 7 in \cite{Vollmann2006} shows that the measured uncertainty of the equivalent width negatively correlates with the equivalent width itself and the flux ratio between the emission lines and the continuum level. In this case, the uncertainties we obtained from the mock sample, which effectively has EW (OII)$\sim$0 \AA, would be a conserved upper limit to the real uncertainty. In this work, we use a 4-order polynomial to characterise $\sigma$(EW) as a function of S/N, as indicated by the boundary of the shaded region in Fig.\ref{fig:mock_EW}. For each Wall galaxy, we estimate the S/N of its spectrum near restframe 3727 {\AA} using the RMS variation in the residual, and assign a corresponding uncertainty $\sigma$(EW) to characterise the uncertainty of its EW(OII) measurement.

\section{Influence of model assumptions \& uncertainties}
\label{app:uncertainties}
\label{subsec:uncertainties}
This study employs a physically motivated yet simplified model to fit the observed spectra and SEDs that allows to infer the star formation and chemical evolution histories of our sample galaxies. As with any model, its underlying assumptions influence the results. Here, we discuss the main simplifications of our model, the implications for our results and the potential uncertainties associated with our findings.

A first potential limitation of our model is that we adopt a single exponentially decaying gas infall model. A more complex star formation history, including multiple epochs of star formation activity, is challenging to constrain due to the limited time resolution achievable in spectral analysis \citep{Zibetti2024}, particularly since we are dealing with stacked spectra, which mix the individual SFHs of different galaxies. At the same time, this parametrisation, still widely used in SFH \citep{BC03} and chemical evolution studies \citep[e.g.][]{Spitoni2017,Lian2018mzr}, succeeds in capturing different star formation scenarios. 
Indeed, in our work, we find that for low-mass galaxies, the gas infall timescale is very long ($\approx$ 10 Gyr), making the gas infall process effectively resembling constant gas accretion. In contrast, for high-mass galaxies, we observe a clearer decline in their gas infall over a shorter time, aligning well with investigations based on non- parametric approaches \citep{Camps-Farina2023}.

We assume that the time-dependent fraction of gas turning into stars during the star formation process is described by a linear Schmidt law \citep{Schmidt1959}, where $S$ is the star-formation efficiency (see equation \eqref{eq:schmidtlaw}). 
We are aware that the SFR correlates more tightly with the gas surface density \citep{Kennicutt1998ApJ} than with total gas mass content. 
In our case, however, modelling the gas surface density requires adopting a disk model and estimating the disk radius, something we cannot do as we are analysing stacked spectra from a sample of galaxies with diverse morphologies. 
We also assume a typical value of $S=1~{\rm Gyr}^{-1}$ \citep{Spitoni2017, Zhou2022} for our analysis, while star formation efficiency (SFE) may, in principle, vary among different galaxies. Our tests indicate that varying the SFE between a range of 0.3 Gyr$^{-1}$ to 2 Gyr$^{-1}$, which encompasses the typical range of SFE values estimated from the extended Kennicutt-Schmidt law \citep{Shi2011,Zhou2022}, results in only minor changes in individual spectral fitting results, while the global trends—i.e., systematic differences between galaxies in different environments—results from different SFE settings, remain essentially unchanged.

We adopt an instantaneous mixing approximation to derive the evolution of the galaxy's metal content, meaning that the gas in a galaxy is always well mixed during its evolution, but in real galaxies the release of different elements, such as $\alpha$-elements from core-collapse supernovae and iron-peak elements from type Ia supernovae \citep{Worthey1994}, occurs on different timescales. The quality of our current data prevents us from modelling these processes in detail. However we are working with relatively young galaxies at $z\sim0.7$, so we do not expect their evolution to differ significantly from the results of an instantaneous approximation.

Another strong assumption adopted is the parametrization of gas removal, that is modelled as proportional to star the formation activity, via a mass-loading factor $\lambda$ primarily describing supernova-driven outflows.   Feedback from AGN and external processes like tidal stripping may operate on different timescales and fully incorporating their descriptions is quite challenging. A simpler alternative approach could be to consider time-dependent $\lambda$ coefficient  as explored in some previous studies \citep[e.g.][]{Lian2018mzr, Zhou2022}. However, the limitations  imposed by the available data quality constrain us to adopting a simple, time-invariant mass-loading factor, which is also commonly employed in several studies \citep[e.g.][]{Spitoni2017, Belfiore2019bathtub, Beverage2021}. In this framework, a large value of $\lambda$ indicates  strong gas removal process, but we cannot directly distinguish between its physical origins. Luckily, the observed mass and environmental dependence of the gas removal strength provides additional clues about the underlying physical processes, as discussed in Section 4.3.

Our model does not contain any explicit merger prescriptions, which prevents to distinguish between stars formed in-situ and those formed ex-situ.  Since we are fitting stacked spectra of galaxies, we primarily trace their dominant stellar component, making it difficult to discern stars formed ex-situ. In fact, even in chemical evolution modelling of the Milky Way, where individual stars can be resolved, most current scenarios \citep[e.g.][]{spitoni2024} do not explicitly account for merger-originated stars (i.e. stars with the same age but different metallicities). Nevertheless, these models still provide a reasonable fit to the major chemical components of the Milky Way. Similarly, given that our sample consists of disk star-forming galaxies, recent mergers are expected to have minimal impact.

Finally, some degeneracy between different parameters of the model are unavoidable. As discussed in \cite{Zhou2022}, fixing yield and star formation efficiency mitigates the degeneracy between $\lambda$ and these parameters. However, since we allow only a single epoch  gas infall, some level of degeneracy between the gas infall look-back time $t_{0}$ and the gas infall timescale $\tau$ is expected, as older, extended star formation episodes can mimic younger, short-duration ones. Nevertheless, environmental variations in best-fit parameters found in our analysis occur perpendicular to this degeneracy, reinforcing the robustness of our conclusions. Bootstrapping tests further confirm that parameter degeneracies do not drive the observed trends.
 We can therefore conclude that the degeneracy in model parameters does not impact the systematic variations observed between different environments.

In summary, this work adopts a simple yet comprehensive model to derive physically motivated SFHs and ChEHs that can be used to fit the observed spectra. While many of the assumptions are kept simple due to the limitations in data quality, the results still provide valuable insights into the physical processes, including gas inflow and gas removal, that have shaped the evolution of our sample galaxies. We anticipate that the high-quality spectral data expected in future surveys will enable a more detailed analysis of these physical processes and offer a deeper understanding of how and when the environment has influenced galaxy evolution.
\end{appendix}

\end{document}